\documentclass[aps,twocolumn,showpacs]{revtex4}
\usepackage{graphicx}
\usepackage{epsfig}
\usepackage{epstopdf}
\usepackage{amsfonts}
\usepackage{amssymb}
\usepackage{amsbsy}
\usepackage{amsmath}
\usepackage{mathrsfs}
\usepackage{latexsym}
\usepackage{natbib}
\setcitestyle{square, comma, numbers,sort&compress, super}
\usepackage{bm}
\usepackage{ifpdf}

\usepackage{color}
\usepackage{braket}
\usepackage{slashed}
\usepackage{pgfplots}

\DeclareMathOperator{\arcsinh}{asinh}

\newcommand{\x}{\mathbf{x}}

\def\k{\mathrm{k}}

\newcommand{\kfb}{\bar{k}_F} 
 
\def\kfb{\bar{k}_F}

\def\mt{\bar{m}}

\def\muimb{\bar{\mu}_{m}}

\begin{document}

\title{The methods of thermal field theory for degenerate quantum plasmas
in astrophysical compact objects}

\author{Golam Mortuza Hossain}
\email{ghossain@iiserkol.ac.in}

\author{Susobhan Mandal}
\email{sm17rs045@iiserkol.ac.in}

\affiliation{ Department of Physical Sciences, 
Indian Institute of Science Education and Research Kolkata,
Mohanpur - 741 246, WB, India }
 

\date{\today}

\begin{abstract}

In the study of degenerate plasmas contained within compact astrophysical 
objects, both special relativity and general relativity play important roles. 
After reviewing the existing treatment in the literature, here we employ the 
methods of relativistic thermal quantum field theory to compute the equation of 
states of degenerate matter for compact astrophysical objects such as the 
white dwarfs and the neutron stars. In particular, we compute the equation of 
states that include leading order corrections due to the finite temperature, the 
fine-structure constant as well as the effect of gravitational time dilation. We 
show that the fine-structure constant correction remains well-defined even in 
the non-relativistic regime in contrast to the existing treatment in the 
literature.

\end{abstract}

\maketitle

\section{Introduction}\label{Introduction}

It is not often that one sees a path-breaking discovery like the observation of 
gravitational waves that pushes the boundary of the fundamental physics into a 
new frontier. The existence of such waves arising from the fluctuations of the 
very fabric of the spacetime, was predicted from Einstein's general relativity 
nearly a century ago. However, the experimental detection of these waves has 
required an unprecedented level of progress in the instrumentation techniques 
and data analysis methods. The very first observation that was made by the 
gravitational wave detectors, is thought to have originated from the merger 
event of two black holes \cite{abbott2016observation}. Subsequently, these 
gravitational wave detectors have detected signals from the merger events of 
binary neutron stars. Remarkably, this event was also accompanied with the 
electromagnetic signals which were captured by multiple telescopes over 
a wide electromagnetic spectrum \cite{abbott2017gw170817}. This unprecedented 
observation of a given event through both electromagnetic waves and 
gravitational waves has led to an era of the so-called multi-messenger 
astronomy. Apart from the already operational ground-based gravitational wave 
detectors, the future space-based gravitational wave detectors, such as LISA 
\cite{shaddock2008space}, are also under active consideration. These detectors 
will potentially detect gravitational wave signals from the merger events 
involving the white dwarf stars.

The common aspect of these gravitational waves detectors is that they 
are primarily designed to observe gravitational waves which essentially 
originate from the coalescence events involving various \emph{massive} yet 
\emph{compact} astrophysical objects. These compact astrophysical objects are 
primarily the black holes, the neutron stars and the white dwarfs. Among these 
astrophysical objects, the most compact objects are the black holes whose 
interior spacetime are screened by an event horizon from where even the light 
cannot escape. This property of the event horizon essentially limits the 
possibility of understanding the interior black hole spacetime even in 
principle. The next most compact astrophysical objects are the neutron stars. 
For example, the neutron star in the well-known Hulse -Taylor binary pulsar 
system has a mass of around $1.4 M_{\odot}$ whereas it has a radius of only 
about ten kilometer or so \cite{hulse1975discovery, taylor1989further}. Such a 
compact object would require its interior energy density to be order of the 
nuclear density and the interior matter to be in some form of a nuclear matter 
plasma. The least compact objects, among the astrophysical objects under 
consideration, are the white dwarf stars or simply the white dwarfs. For 
example, the famous white dwarf Sirius B whose mass is almost equal to the mass 
of Sun but has a radius which is even less than the radius of the planet Earth. 
A simple calculation would show that even for the white dwarfs interior 
matter density is so high that on the average an electron has lesser space 
inside a white dwarf than the space which is required for the electron to be in 
an atomic orbital. Clearly, inside the white dwarfs the matter can only be in 
the form of a plasma as the atoms themselves are smashed by the enormous 
gravitational pull. However, the electrons in the case of white dwarfs or the 
neutrons in the case of neutron stars are fermions and due to Pauli's exclusion 
principle, these fermions resist being compressed to an extreme density. The 
resultant pressure of quantum mechanical origin that counters the gravitational 
pull to make these stars stable, is usually known as the \emph{degeneracy} 
pressure.

Clearly, in the view of gravitational wave observations with an unprecedented 
precision, it has become a necessity to understand the physics of these compact 
astrophysical objects in a much more detailed manner. This in turn demands one 
to study the properties of the degenerate quantum plasmas that are present 
within these objects, using the best possible theoretical framework that are 
known today. 
Here we describe the degenerate quantum plasmas inside 
the self-gravitating stars by using the methods of quantum field theory in 
curved spacetime. In particular, we consider Einstein's equation of 
general relativity which is of the form $G_{\mu\nu} = 8\pi G~
\langle\hat{T}_{\mu\nu}\rangle$ where $\langle .\rangle$ denotes the thermal 
expectation value. In this formulation, the matter degrees of freedom,
represented by the stress-energy tensor ${T}_{\mu\nu}$, are described by the 
quantum theory. On the other hand, the spacetime curvature, represented by the 
Einstein tensor $G_{\mu\nu}$, is described by the classical theory. We may 
emphasize here that we do not use any quantum theory of gravity. Firstly, the 
quest for a quantum theory of gravity is still an area of active research. 
Secondly, for compact stars such as the white dwarfs or the neutron stars,
the quantum gravity corrections are expected to be rather negligible, as the 
energy scales associated with these stars are extremely small compared to the 
Planck scale, the presumed scale of quantum gravity.

In this review, we begin by considering a system of degenerate quantum plasmas 
in the background of Minkowski spacetime. Such descriptions are directly 
relevant for describing the equation of state (EOS) corresponding to the plasma 
of degenerate electrons which are present within the white dwarfs. For these 
stars, the effects of general relativity are relatively smaller. In order to 
describe degenerate plasma inside the white dwarfs one also needs to consider 
electromagnetic interaction between the electrons themselves and also between 
the electrons and the positively charged nuclei. Subsequently, we consider the 
effects of general relativity on the equation of states of these degenerate 
plasmas. In particular, we study the equation of state of the degenerate plasmas 
which are located within a spherically symmetric curved spacetime. We then 
consider the effects of these corrections on the mass-radius relations of the 
white dwarfs. On the other hand, due to the limitations in understanding of the 
dense nuclear matter physics, the understanding of the degenerate plasmas 
present within the neutron stars are still an active area of ongoing research. 
Nevertheless, the simplest form of degenerate plasma inside the neutron star can 
be viewed as an ensemble of degenerate neutrons. These neutrons being neutral 
particles they do not interact electromagnetically. The corresponding equation 
of state can be obtained from the equation of state of degenerate electrons by 
suitably mapping the parameters and turning off the electromagnetic 
interactions. Thereafter, we briefly discuss about a widely used realistic 
model, the so-called the $\sigma-\omega$ model, to describe the degenerate 
nuclear matter present within the neutron stars.

\section{Degenerate plasma in flat spacetime}\label{DegenerateFlat}

In the study of degenerate matter within the white dwarfs, as pioneered by 
Chandrasekhar \cite{chandrasekhar1931maximum, chandrasekhar1935highly}, the 
degenerate electrons are treated as free particles which follow the Fermi-Dirac 
statistics. The effect of electromagnetic interaction on the degenerate matter 
in the white dwarfs, by the means of `classical' Coulomb energy, was first 
considered by Frenkel \cite{frenkel1928}, and was followed up by Kothari 
\cite{kothari1938theory}, Auluck and Mathur \cite{auluck1959electrostatic}. 
However, a  more accurate study on various corrections to the equation of state 
due to the so-called Coulomb effects was done by Salpeter 
\cite{salpeter1961energy}. The implications of these corrections on the 
mass-radius relation of the white dwarfs were carried out by Hamada, Salpeter 
\cite{hamada1961models} and Nauenberg \cite{nauenberg1972analytic}.

One may classify the total Coulomb effects that are considered by Salpeter into 
following broad components (see also 
\cite{koester1990physics,book:Shapiro.Teukolsky}). (a) The `classical' Coulomb 
energy includes the electrostatic energy of uniformly distributed degenerate 
electrons within the Wigner-Seitz cells, each surrounding a positively charged 
nucleus within a rigid lattice. It includes the electron-nuclei interaction and 
the self-interaction of electrons. (b) The Thomas-Fermi correction arises  due 
to the radial variation of electron density within a Wigner-Seitz cell. (c) The 
`exchange energy' and the `correlation energy' arise due to the transverse 
interactions between two electrons, essentially due to the Lorentz force 
between them apart from its electrostatic component which is already included in 
(a). We may mention here that the Thomas-Fermi model is a non-relativistic model 
and relativistic corrections to it have been considered for the white dwarfs in 
Ref. \cite{rotondo:PhysRevC.83.045805,rotondo:PhysRevD.84.084007, 
rotondo:PhysRevC.89.015801}.

The special relativity plays a key role in the white dwarf physics. In 
particular, the existence of the Chandrasekhar upper mass limit for the white 
dwarfs arises essentially due to the special relativity which demands that the 
physical results should be invariant under the Lorentz transformations. 
Unfortunately, the methods employed in computing the Coulomb effects to the 
equation of state of the white dwarfs use electrostatic considerations which are 
non-relativistic \emph{ab initio}. Therefore, from the fundamental point of view 
these computations should be viewed as an approximation of the corrections that 
one would expect from a Lorentz invariant computation. Further, these 
computations are usually performed at the zero temperature 
\cite{salpeter1961energy} (however see \cite{kovetz1970thermodynamics, 
shaviv1972thermodynamics, Fantoni:2017mfs}). 
the future detection of low-frequency gravitational waves from the extreme 
mass-ratio merger of a black hole and a white dwarf could determine the equation 
of state of the degenerate matter within the white dwarfs with very high 
accuracy \cite{Han:2017kre}. Such an accuracy could probe the extent of Coulomb 
effects and hence provides additional motivation to revisit the corrections to 
the equation of state of the white dwarfs.

In order to compute the equation of state for the white dwarfs, a natural arena 
which respects Lorentz symmetry, is provided by the relativistic thermal 
quantum field theory, also referred to as the finite temperature 
relativistic quantum field theory. Following the pioneering work of Matsubara 
\cite{matsubara1955new}, the techniques of finite-temperature quantum field 
theory was employed in the context of \emph{quantum electrodynamics} (QED) by 
Akhiezer and Peletminskii \cite{akhiezer1960use}, and later by Freedman and 
McLerran \cite{PhysRevD.16.1147}, to compute the ground state energy of the 
relativistic electron gas that includes corrections due to the fine-structure 
constant. However, these treatments are insufficient to describe the degenerate 
matter in the white dwarfs as they do not describe the dominant interaction, as 
seen in non-relativistic computations,  between the degenerate electrons and 
positively charged heavier nuclei. Therefore, in this review article, to 
describe the degenerate matter within the white dwarfs using the framework of 
finite temperature quantum field theory, we consider an additional interaction 
between the electrons and positively charged nuclei, described by a Lorentz 
invariant action, together with the quantum electrodynamics.

\subsection{The scale of degeneracy}

In order to understand the associated scales of the degenerate matter, let us 
consider the white dwarf star Sirius B which has observational mass $M = 1.0 
~M_{\odot}$, radius $R = 0.008~R_{\odot}$ and the effective temperature $T = 
25922~\mathrm{K}$ \cite{joyce:2018mnras481}. Therefore, its mass density is 
$\rho \approx 2.8 \times 10^6 ~g/cm^3$. In \emph{natural units} that we follow 
here (\emph{i.e.} speed of light $c$ and Planck constant $\hbar$ are set to 
unity), a fully degenerate core implies that the electron density is $n_e 
\approx 6.4 \times 10^{15} ~\mathrm{(eV)^3}$. The corresponding Fermi momentum 
is $k_F \simeq (3\pi^2 n_e)^{1/3} \approx 5.7 \times 10^5 ~\mathrm{eV}$. The 
associated temperature scale of the white dwarfs $\beta^{-1} \equiv k_B T = 
2.2~\mathrm{eV}$ then leads to a dimensionless parameter 
\begin{equation}\label{BetaKfRatio}
\beta k_F \approx 2.6 \times 10^5  ~,
\end{equation}
which plays an important role in characterizing the degenerate matter within 
the white dwarf stars.

\subsection{Dirac spinor field}

In order to compute the equation of state of the degenerate matter, we consider 
the spacetime within a white dwarf star to be described by the Minkowski metric 
$\eta_{\mu\nu} = diag(-1,1,1,1)$ \emph{i.e.} for time being we ignore the 
corrections from the \emph{general relativity} (as also done by Salpeter 
\cite{salpeter1961energy}). Nevertheless, we shall be considering these
corrections arising from the general relativity in a later section. The 
degenerate electrons are fermionic degrees of freedom and are represented by the 
Dirac spinor field $\psi$ along with the action
\begin{equation}\label{FermionAction}
S_{\psi} = \int d^{4}x \mathcal{L}_{\psi}  = -\int d^{4}x \sqrt{-\eta} 
~ \overline{\psi}[i \gamma^{\mu} \partial_{\mu} + m]\psi  ~,
\end{equation}
where $\eta = det(\eta_{\mu\nu}) = -1$ and $m$ is the mass of the field. The 
Dirac matrices $\gamma^{\mu}$ satisfy the anti-commutation relation
\begin{equation}\label{DiracMatrices}
\{\gamma^{\mu},\gamma^{\nu}\} = - 2 \eta^{\mu\nu} \mathbb{I} ~.
\end{equation}
The minus sign in front of $\eta^{\mu\nu}$ in the Eq. (\ref{DiracMatrices}) is 
chosen such that for given metric signature, the Dirac matrices satisfy the 
usual relations $(\gamma^0)^2 = \mathbb{I}$ and $(\gamma^k)^2 = -\mathbb{I}$ 
for $k=1,2,3$. In Dirac representation, these matrices can be expressed as 
\begin{equation}\label{DiracMatricesUsingPauli}
\gamma^0 = \begin{pmatrix} \mathbb{I} & 0 \\ 0 & -\mathbb{I} \end{pmatrix} ~~, 
~~ \gamma^k = \begin{pmatrix} 0 & \sigma^k \\ -\sigma^k & 0 \end{pmatrix} ~,
\end{equation}
where Pauli matrices $\sigma^k$ are given by
\begin{equation}\label{PauliMatrices}
\sigma^1 = \begin{pmatrix} 0 & 1 \\ 1 & 0 \end{pmatrix} ~~,~~
\sigma^2 = \begin{pmatrix} 0 & -i \\ i & 0 \end{pmatrix} ~~,~~
\sigma^3 = \begin{pmatrix} 1 & 0 \\ 0 & -1 \end{pmatrix} ~.
\end{equation}

\subsection{Gauge field}\label{Gauge field}

The electromagnetic interaction between the fermions are mediated by the gauge 
fields $A_{\mu}$ whose free dynamics is governed by the Maxwell action
\begin{equation}\label{MaxwellAction}
S_{A} = \int d^4x \mathcal{L}_{A} = \int d^{4}x ~ \left[ 
-\frac{1}{4} F_{\mu\nu} F^{\mu\nu} \right]  ~,
\end{equation}
where the field strength $F_{\mu\nu} = \partial_{\mu}A_{\nu} - 
\partial_{\nu}A_{\mu}$.

\subsection{Field interactions}

The degenerate electrons within a white dwarf experience two kinds of 
interactions, namely the self-interaction between the electrons and the 
interaction between electrons and the positively charge nuclei. The 
self-interaction between the electrons are mediated by gauge fields $A_{\mu}$ 
and governed by the interaction term of the quantum electrodynamics 
\begin{equation}\label{InterctionAction}
S_{I}^{-} = \int d^4x \mathcal{L}_{I}^{-} = \int d^{4}x ~ 
\bar{\psi}[ e ~ \gamma^{\mu} A_{\mu} ]\psi ~,
\end{equation}
where the parameter $e$ is the dimensionless coupling constant. 

We may recall that the conserved 4-current corresponding to the action 
(\ref{FermionAction}) is given by $j^{\mu} = \bar{\psi} \gamma^{\mu} \psi$ 
which represents the contribution of the electrons. Similarly, we may consider 
a 
background 4-current, say $J^{\mu}$, to represent the contributions from the 
nuclei. Therefore, we may model the attractive interaction between the 
electrons and the positively charged nuclei by a Lorentz invariant action 
containing the \emph{current-current} interaction 
\begin{equation}\label{InteractionActionNucleon}
S_{I}^{+} = \int d^4x \mathcal{L}_{I}^{+} = \int d^{4}x  ~ 
[ - Z e^2 d^{2} ~ J_{\mu} \bar{\psi}\gamma^{\mu}\psi] ~,
\end{equation}
where the coupling constant contains the term $- Z e^2 $ which signifies the 
strength of the attractive interactions between an electron and a positively 
charged nucleus with atomic number $Z$. The parameter $d$ is introduced in 
order to make the action (\ref{InteractionActionNucleon}) dimensionless. It has 
the dimension of length and and it represents an effective length scale 
associated with the current-current interaction.

Therefore, the total action that describes the dynamics of the degenerate 
electrons within a white dwarf can be written as
\begin{equation}\label{TotalMatterAction}
S =  S_{\psi} + S_A + S_{I} =  S_{QED} + S_{I}^{+}  ~,
\end{equation}
where $S_{I} = S_{I}^{-} + S_{I}^{+}$. The inclusion of the additional 
interaction term (\ref{InteractionActionNucleon}) preserves the symmetry of the 
action of quantum electrodynamics $S_{QED}$. In other words, apart from being 
Lorentz invariant, the total action (\ref{TotalMatterAction}) is also invariant 
under local U(1) gauge transformations $A_{\mu}\rightarrow 
A_{\mu}-\frac{1}{e}\partial_{\mu}\alpha(x)$ and $\psi(x) \rightarrow 
e^{i\alpha(x)}\psi(x)$ with $\alpha(x)$ being an arbitrary function. Given the 
coupling constant $e$ is small, we can study the interacting theory by using 
perturbative techniques.

\subsection{Partition function}

In a spherically symmetric star, the pressure and the energy density both vary 
along the radial direction. On the other hand, in order to apply the techniques 
of finite temperature quantum field theory we need to consider a spatial region 
which is in thermal equilibrium at a given temperature $T$. Within such a
region thermodynamical quantities such as the pressure and the density are 
uniform. Therefore, in order to deal with both these aspects we consider here a 
finite spatial box at a given radial coordinate. The box is assumed to be 
sufficiently small so that the pressure and the density remain uniform within 
the box yet it is sufficiently large to contain enough degrees of freedom to 
achieve required thermodynamical equilibrium. 
The corresponding partition function, describing the degrees of freedom within 
the box, can be expressed as
\begin{equation}\label{PartionFunction0}
\mathcal{Z} = Tr \left[e^{ -\beta (\hat{H} - \mu \hat{Q})} \right] ~,
\end{equation}
where $\beta = 1/k_B T$ with $k_B$ being the Boltzmann constant, $\mu$ refers 
to chemical potential and $Q$ is the conserved charge of the system. The 
Hamiltonian operator $\hat{H}$ represents the matter fields described by the 
action (\ref{TotalMatterAction}). The trace operation is carried out over the 
degrees of freedom contained within the specified spatial region.

\subsubsection{Partition function for free fermions}

The action (\ref{FermionAction}) of free spinor field is invariant under a 
global U(1) gauge transformations $\psi(x) \rightarrow e^{i\alpha}\psi(x)$ 
where $\alpha$ is an arbitrary constant. Consequently, there exists an 
associated conserved current $j^{\mu} = \bar{\psi} \gamma^{\mu} \psi$ such 
that $\partial_{\mu} j^{\mu} = 0$. Then the corresponding conserved charge
can be expressed as $Q = \int d^{3}\x~ j^0(x) = \int d^{3}\x 
~\bar{\psi}\gamma^{0}\psi$. Additionally, the conjugate field momentum 
corresponding to the spinor field $\psi(x)$ can be expressed as $\pi(x) = 
\frac{\partial\mathcal{L}_{\psi}}{\partial(\partial_{0}\psi)} = - i \bar{\psi} 
\gamma^{0} = - i \psi^{\dagger}$. Therefore, the partition function which 
contains contributions only from free spinor field can be expressed using 
path integral method \cite{book:kapusta} as
\begin{equation}\label{SpinorPartitionFunction}
\mathcal{Z}_{\psi} = \int \mathcal{D}\bar{\psi} \mathcal{D}\psi 
 ~ e^{- S_{\psi}^{\beta}} ~,
\end{equation}
where
\begin{equation}\label{PartitionFunctionSBE}
S_{\psi}^{\beta} = \int_0^{\beta}d\tau \int d^3\x~ 
\left[\mathcal{L}_{\psi}^{E} - 
\mu \bar{\psi}(\tau,\x) \gamma^0\psi(\tau,\x) \right] ~.
\end{equation}
The Euclidean Lagrangian density $\mathcal{L}_{\psi}^{E}$ is obtained by 
substituting $t\to -i\tau$ in Lagrangian density $\mathcal{L}_{\psi}$ and can 
be expressed as 
\begin{equation}\label{EuclideanLagrangianDensity}
\mathcal{L}_{\psi}^{E} = \bar{\psi}(\tau,\x)[- \gamma^{0} \partial_{\tau} + 
i \gamma^{k} \partial_{k} + m]\psi(\tau,\x)  ~,
\end{equation}
where $k=1,2,3$. In the functional integral (\ref{SpinorPartitionFunction}), 
the spinor field is subject to \emph{anti-periodic} boundary conditions given by
\begin{equation}\label{FermionicBoundaryCondition}
\psi(\tau,\x) = -\psi(\tau+\beta,\x)   ~~;~~
\bar{\psi}(\tau,\x) = -\bar{\psi}(\tau+\beta,\x) ~.
\end{equation}
It is convenient to express the partition function using the Matsubara 
frequencies and wave-vector by transforming the field in Fourier domain as
\begin{equation}\label{FermionicFourier}
\psi(\tau,\x) = \frac{1}{\sqrt{V}} \sum_{n,\k} ~e^{-i(\omega_n\tau + \k\cdot\x)} 
\tilde{\psi}(n,\k)  ~,
\end{equation}
where $V$ denotes the spatial volume of the box. The Eq. 
(\ref{FermionicBoundaryCondition}) then implies that the Matsubara frequencies 
are
\begin{equation}\label{FermionicOmegan}
\omega_n = (2n+1) \pi ~\beta^{-1}  ~,
\end{equation}
where $n$ is an integer. The Eqs. (\ref{PartitionFunctionSBE}) and 
(\ref{FermionicFourier}) then lead to
\begin{equation}\label{PartitionFunctionSBEFourier}
S_{\psi}^{\beta} = \sum_{n,\k} ~\bar{\tilde{\psi}}~\beta
\left[ \slashed{p} + m \right]  \tilde{\psi} ~,
\end{equation}
where $p_{\mu} = (p_0,\vec{p}) = (i\omega_n-\mu,\k)$ and $\slashed{p} = 
\gamma^{\mu} p_{\mu}$. The spinor fields $\psi$ and $\bar{\psi}$ satisfy the 
same algebra as the Grassmann variables. 
Using the Dirac representation of the gamma matrices and the result of Gaussian 
integral over Grassmann variables one would get
$
\ln\mathcal{Z}_{\psi} = 2 \sum_{\k} [ \beta \omega
+ \ln (1 + e^{-\beta(\omega-\mu)}) + 
\ln (1 + e^{-\beta(\omega+\mu)} ) ].
$
Here the factor of 2 denotes the spin-degeneracy of the fermions and the first 
term inside the bracket corresponds to the zero-point energy which is formally 
divergent. The second and third terms here correspond to the contributions 
coming from the particles and the anti-particles respectively. Given $\beta \mu 
\gg 1$ for a degenerate system, the anti-particle contributions are 
exponentially suppressed. By disregarding the contributions from the 
anti-particles and the formally divergent terms, we can evaluate the partition 
function as
\begin{equation}\label{LogPartitionFunctionFermionFinal}
\ln\mathcal{Z}_{\psi} = \frac{\beta V}{24\pi^2} 
\left[2\mu k_F^3 - 3m^2 \kfb^2 + \frac{48 \mu k_F}{\beta^{2}} \right]  ~,
\end{equation}
where 
\begin{equation}\label{kfbDef}
\kfb^2 \equiv \mu k_F  - m^2 \ln \left(\frac{\mu + k_F }{m}\right) ~.
\end{equation}
In the Eq. (\ref{LogPartitionFunctionFermionFinal}), we have also ignored 
higher order temperature corrections which are at least 
$\mathcal{O}((\beta\mu)^{-2})$.

\subsubsection{Partition function for photons}

Due to the gauge symmetry  $A_{\mu}(x)$ and $A'_{\mu}(x) = A_{\mu}(x) 
-\tfrac{1}{e} \partial_{\mu} \alpha(x)$ represent the same physical 
configuration. Therefore, in order to avoid over-counting in evaluating the 
partition function using functional integral methods, it is convenient to 
introduce the Faddeev-Popov ghost fields $C$ and $\bar{C}$ 
\cite{book:17045,book:16435}. These Grassmann-valued fields effectively cancel 
the contributions from two gauge degrees of freedom. Therefore, the thermal 
partition function containing contributions from the physical photons can be 
expressed as 
\begin{equation}\label{PhotonPartitionFunction}
\mathcal{Z}_{A} = \int \left(\mathcal{D}A_{\mu} ~e^{-S_{A}^{\beta}} \right)
\left(\mathcal{D}\bar{C}~\mathcal{D}C ~e^{-S_{C}^{\beta}} \right)
\equiv \mathcal{Z}_{A'} \mathcal{Z}_{C} ~,
\end{equation}
where $S_{A}^{\beta} = -\int_0^{\beta}d\tau \int d^3\x ~
[-\tfrac{1}{4} F_{\mu\nu} F^{\mu\nu} -\tfrac{1}{2\xi} (\partial_{\mu} 
A^{\mu})^2]_{|t=-i\tau}$ with gauge-fixing parameter $\xi$ and $S_{C}^{\beta} 
= \int_0^{\beta}d\tau \int d^3\x ~ [\partial^{\mu}\bar{C} \partial_{\mu} 
C]_{|t=-i\tau}$. Unlike the spinor field, both $A_{\mu}(x)$ and $C(x)$ fields 
are subject to the \emph{periodic} boundary conditions 
\begin{equation}\label{GaugeFieldBoundaryCondition}
A_{\mu}(\tau,\x) = A_{\mu}(\tau+\beta,\x)   ~~;~~
C(\tau,\x) = C(\tau+\beta,\x) ~.
\end{equation}
As earlier, we evaluate the partition function in Fourier domain
by transforming the field as
\begin{equation}\label{GaugeFieldFourier}
A_{\mu}(\tau,\x) = \sqrt{\frac{\beta}{V}} ~ \sum_{n,\k} 
~e^{-i(\omega_n\tau + \k\cdot\x)} \tilde{A_{\mu}}(n,\k)  ~.
\end{equation}
The definition (\ref{GaugeFieldFourier}) ensures that the Fourier modes 
$\tilde{A_{\mu}}(n,\k)$ are dimensionless and the Eq. 
(\ref{GaugeFieldBoundaryCondition}) implies $\omega_n = 2n\pi 
\beta^{-1}$ with $n$ being an integer. By choosing Feynman gauge $\xi=1$ and 
dropping the boundary terms, we can express $S_{A}^{\beta}$ as
\begin{equation}\label{PartitionFunctionSBAFourier}
S_{A}^{\beta} = \sum_{n,\k} \bar{\tilde{A_{\mu}}}
\left[ \frac{1}{2}\beta^2(\omega_n^2 + \k^2) \right] \tilde{A_{\nu}} 
\delta^{\mu\nu} ~.
\end{equation}
Using the identity for Riemann integrals $\int x_1 \dots x_N e^{-x_i D_{ij} 
x_j} = \pi^{N/2} (det(D))^{-1/2}$, we can evaluate the contributions from the 
gauge field by
\begin{equation}\label{LogPartitionFunctionOnlyGaugeField}
\ln\mathcal{Z}_{A'} = -2 \sum_{n,\k} 
\ln\left[\beta^2 (\omega_n^2 +\k^2) \right] ~
\end{equation}
where the gauge field is Wick rotated as $\tilde{A_0} \to i \tilde{A_0}$ to 
make the integral convergent. Similarly, one can define the Fourier modes of 
the ghost field as
\begin{equation}\label{GhostFieldFourier}
C(\tau,\x) = \sqrt{\frac{\beta}{V}} ~ \sum_{n,\k} 
~e^{-i(\omega_n\tau + \k\cdot\x)} \tilde{C}(n,\k)  ~,
\end{equation}
where the modes $\tilde{C}(n,\k)$ are again dimensionless and $\omega_n = 
2n\pi \beta^{-1}$ with $n$ being an integer. By dropping the boundary terms, we 
can the express $S_{C}^{\beta}$ as
\begin{equation}\label{PartitionFunctionSBCFourier}
S_{C}^{\beta} = \sum_{n,\k} \bar{\tilde{C}}~
\left[ \beta^2 (\omega_n^2 + \k^2) \right] \tilde{C} ~.
\end{equation}
The ghost fields $C$ and $\bar{C}$ being Grassmann-valued field, we can use the 
same identity as used for fermions in order to evaluate their contributions as
\begin{equation}\label{LogPartitionFunctionOnlyGhosts}
\ln\mathcal{Z}_{C} =  \sum_{n,\k} 
\ln\left[\beta^2 (\omega_n^2 +\k^2) \right] ~.
\end{equation}
By combining the contributions (\ref{LogPartitionFunctionOnlyGaugeField}) and 
(\ref{LogPartitionFunctionOnlyGhosts}) one can write the partition function for 
the physical photons as
\begin{equation}\label{LogPartitionFunctionGaugeFieldFinal}
\ln\mathcal{Z}_{A} = \frac{V\pi^2}{45 \beta^3} ~.
\end{equation}

\subsubsection{Contributions from the interactions}

Including both kinds of interaction for the degenerate electrons, we can 
express total partition function as
\begin{equation}\label{TotalPartitionFunctionIntegral}
\mathcal{Z} = \int \mathcal{D}\bar{\psi} \mathcal{D}\psi
\mathcal{D}A_{\mu} \mathcal{D}\bar{C}~\mathcal{D}C 
~ e^{-(S_{\psi}^{\beta} + S_{A}^{\beta} + S_{C}^{\beta} + S_{I}^{\beta})} ~,
\end{equation}
where $S_{I}^{\beta} = S_{-}^{\beta} + S_{+}^{\beta}$ with
$S_{-}^{\beta} = \int_0^{\beta}d\tau \int d^3\x ~ 
[\mathcal{L}_{I}^{-}]_{|t=i\tau}$ and 
$S_{+}^{\beta} = \int_0^{\beta}d\tau \int d^3\x ~ 
[\mathcal{L}_{I}^{+}]_{|t=i\tau}$.
Using perturbative method, the total partition function 
(\ref{TotalPartitionFunctionIntegral}) can be expressed as
\begin{equation}\label{LogPartionFunction}
\ln\mathcal{Z} = \ln\mathcal{Z}_{\psi} + \ln\mathcal{Z}_A + 
\ln\mathcal{Z}_{I} ~,
\end{equation}
where the contribution due to the interactions is
\begin{equation}\label{LogInteractionPartitionFunctionFull}
\ln\mathcal{Z}_{I} =  \ln\left(1 + \sum_{l=1}^{\infty} \frac{1}{l!}~ 
\langle {(- S_{I}^{\beta})}^l \rangle \right)  ~.
\end{equation}
Including only the leading order terms we can express the Eq. 
(\ref{LogInteractionPartitionFunctionFull}) as
\begin{equation}\label{LogPartitionFunctionInteraction}
\ln\mathcal{Z}_{I} = \frac{1}{2} \langle (S_{-}^{\beta})^2 \rangle 
- \langle {(S_{+}^{\beta})} \rangle + \mathcal{O}(e^3)  ~,
\end{equation}
where the symbol $\langle.\rangle$ denotes the ensemble average.

\subsubsection{Finite-temperature propagators}

In order to compute $\ln\mathcal{Z}_{I}$, one needs the finite-temperature 
propagators for the spinor field and the Maxwell's field. In particular, the 
finite-temperature propagator for spinor field in real space is defined as
\begin{equation}\label{SpinorPropagatorRealSpace}
\mathcal{G}^{0}(\Delta\tau,\Delta\x) = 
\langle \psi(\tau_1,\x_1)  \overline{\psi}(\tau_2,\x_2)\rangle ~,
\end{equation}
where $\Delta\tau = \tau_1-\tau_2$, $\Delta\x = \x_1-\x_2$. The corresponding 
propagator in Fourier space is defined as
\begin{equation}\label{SpinorPropagatorFourierDef}
\mathcal{G}(\omega_n,\k) = \int_0^{\beta}d\tau \int d^3\x 
~e^{i(\omega_n\tau + \k\cdot\x)} ~\mathcal{G}^{0}(\tau,\x) ~.
\end{equation}
Using the Eq. (\ref{PartitionFunctionSBEFourier}), the propagator
for the free spinor field in Fourier space can be obtained as
\begin{equation}\label{SpinorPropagator}
\mathcal{G}(\omega_n,\k) = \frac{1}{\slashed{p} + m}
= - \frac{\slashed{p} - m}{p^2 + m^2} ~,
\end{equation}
where $p_{\mu} = (p_0,\vec{p}) = (i\omega_n-\mu,\k)$, $\slashed{p} = 
\gamma^{\mu} p_{\mu}$ and $p^2 = \eta^{\mu\nu} p_{\mu} p_{\nu}$. Similarly, the 
finite-temperature propagator for Maxwell's field in real space is defined as 
$\mathcal{D}_{\mu\nu}(\Delta\tau,\Delta\x) = \langle A_{\mu}(\tau_1,\x_1) 
A_{\nu}(\tau_2,\x_2)\rangle$. Following the Eq. 
(\ref{SpinorPropagatorFourierDef}), the propagator for free Maxwell fields in 
Fourier space can be obtained using the Eq. (\ref{PartitionFunctionSBAFourier}) 
as
\begin{equation}\label{PhotonPropagator}
\mathcal{D}_{\mu\nu}^{0}(\omega_n,\k) = \frac{-\eta_{\mu\nu}}{\omega_n^2+\k^2} 
~.
\end{equation}
We may emphasize here that in the Eq. (\ref{SpinorPropagator}), the 
Matsubara frequencies are $\omega_n = (2n+1)\pi \beta^{-1}$ whereas in the Eq. 
(\ref{PhotonPropagator}), they are $\omega_n = 2n\pi \beta^{-1}$ with $n$ being 
an integer.

\subsubsection{Electron-electron interaction}

Using the Eq. (\ref{InterctionAction}), we can express leading order 
contributions due to the self-interaction of the electrons as
\begin{eqnarray}\label{SMinusSquaredAverageRealSpace}
\langle (S_{-}^{\beta})^2 \rangle = - e^2 \int_0^{\beta}d\tau_1 d\tau_2 \int 
d^3\x_1 d^3\x_2 ~ \mathcal{D}_{\mu\nu}(\Delta\tau,\Delta\x) \nonumber \\
\times ~ \mathrm{Tr} \left[ \gamma^{\mu}\mathcal{G}^{0}(\Delta\tau,\Delta\x) 
 \gamma^{\nu}\mathcal{G}^{0}(-\Delta\tau,-\Delta\x) \right]
 ~,~~ 
\end{eqnarray}
where the trace is carried over the Dirac indices. Here we have dropped the 
divergent diagrams that arise from the usage of the Wick's theorem. The Eq. 
(\ref{SMinusSquaredAverageRealSpace}), can be expressed in terms of the  
propagators in Fourier space as
\begin{eqnarray}\label{SMinusSquaredAverageFourierSpace}
\langle (S_{-}^{\beta})^2 \rangle = - \frac{e^2}{\beta V}
\sum_{n_1,n_2,\k_1,\k_2} ~ 
\mathcal{D}_{\mu\nu}(\Delta\omega_{n},\Delta\k) \nonumber \\ \times ~ 
\mathrm{Tr} \left[\gamma^{\mu}\mathcal{G}(\omega_{n_1},\k_1) 
 \gamma^{\nu}\mathcal{G}(\omega_{n_2},\k_2) \right]  ~,~~ 
\end{eqnarray}
where $\Delta\omega_n = \omega_{n_1} - \omega_{n_2}$ and $\Delta\k = \k_1 - 
\k_2$. Using the Eqs. (\ref{SpinorPropagator}, \ref{PhotonPropagator})  
one can simplify the Eq. (\ref{SMinusSquaredAverageFourierSpace}) as
\begin{equation}\label{SMinusSquaredAverageFourierSpace1}
\langle (S_{-}^{\beta})^2 \rangle = - \frac{4e^2}{\beta V}
\sum_{\substack{n_1,n_2\\ \k_1,\k_2}}  ~ 
\frac{4m^2 + 2 p_1 \cdot p_2 }{(p_1^2+m^2)(p_1-p_2)^2(p_2^2+m^2)}  ~,~~ 
\end{equation}
where $p_1=(i\omega_{n_1}-\mu,\k_1)$ and $p_2=(i\omega_{n_2}-\mu,\k_2)$.
Here we have used the trace identities for the Dirac matrices 
$\mathrm{Tr}(\gamma^{\mu}\gamma^{\nu}) = - 4\eta^{\mu\nu}$,
$\mathrm{Tr}(\gamma^{\mu}\gamma^{\nu}\gamma^{\rho}\gamma^{\sigma}) = 
4(\eta^{\mu\nu}\eta^{\rho\sigma}-\eta^{\mu\rho}\eta^{\nu\sigma}+\eta^{\mu\sigma}
\eta^{\nu\rho})$ and the fact that $(p_1 -p_2)_0 = i \Delta \omega_n$. The Eq. 
(\ref{SMinusSquaredAverageFourierSpace1}) can be written in four parts as 
\begin{equation}\label{SMinusSquaredAverageFourierSpace2}
\langle {(S_{-}^{\beta})}^2 \rangle = - \frac{4e^2}{\beta V}
\left( S_1 + S_2 + S_3 + S_4 \right) ~,
\end{equation}
where
\begin{eqnarray}\label{SDefinition}
S_1 &=& \sum_{n_1,n_2,\k_1,\k_2} \frac{1}{(p_1^2+m^2)(p_1-p_2)^2} ~,\\
S_2 &=& \sum_{n_1,n_2,\k_1,\k_2} \frac{1}{(p_2^2+m^2)(p_1-p_2)^2}  ~,\\
S_3 &=& \sum_{n_1,n_2,\k_1,\k_2} \frac{-1}{(p_1^2+m^2)(p_2^2+m^2)}  ~,\\
S_4 &=& \sum_{n_1,n_2,\k_1,\k_2} \frac{2m^2}{(p_1^2+m^2)(p_1-p_2)^2(p_2^2+m^2)} 
~.
\end{eqnarray}
It can be shown that the term $S_4$ is infrared divergent and hence ignored. 
Further, using the symmetries of the expressions, we may note that $S_1 = S_2 = 
I_0~I_1$, $S_3 = -(I_0)^2$ where
\begin{equation}\label{I0I1Definition}
I_0 = \sum_{n,\k} \frac{1}{p^2+m^2}  ~,~
I_1 = \sum_{n,\k} \frac{1}{(p-p_2)^2}  ~.~
\end{equation}
Despite the appearance of $p_2$ in its expression, the evaluated
$I_1$ does not depend on $p_2$ and is given by
\begin{equation}\label{I1Evaluated}
I_1 = \sum_{n,\k} \frac{\beta^2/4}{n^2\pi^2 + (\tfrac{\beta\k}{2})^2} 
= \sum_{\k} \tfrac{\beta}{|\k|} 
(\tfrac{1}{2} + \tfrac{1}{e^{\beta|\k|}-1} )  = 
\frac{V}{12\beta} ~.
\end{equation}
In order to carry out the summation over Matsubara frequencies, we have used 
the 
identity 
\begin{equation}\label{CothIdentity}
\coth z = \sum_{n=-\infty}^{\infty} \frac{z}{n^2\pi^2 + z^2} ~.
\end{equation}
The summation over $\k$ is carried out by converting it to an 
integral as earlier. Subsequently, by using the Riemann zeta function identity 
$\zeta(2) = \int_0^{\infty} dt ~ t/(e^t-1) = \pi^2/6$ and dropping the 
divergent 
part, we have expressed the finite part of $I_1$.  In order to evaluate $I_0$ 
we 
can express it as
\begin{equation}\label{I0Evaluated1}
I_0 = \sum_{n,\k} \frac{1}{2\omega} 
\left[\frac{1}{p_0 + \omega} + \frac{1}{-p_0 + \omega} \right]
\equiv I_0^{+} + I_0^{-}   ~,
\end{equation}
where $p_0 = -i(2n+1) \pi ~\beta^{-1}+\mu$ and
\begin{equation}\label{I0Evaluated0}
I_0^{\pm} = \sum_{n,\k} \frac{\beta}{4\omega}
\left( \frac{z_{\pm}}{n^2\pi^2 + (z_{\pm})^2}  \right) ~,
\end{equation}
with $z_{\pm} = \tfrac{1}{2}\{\beta(\omega \pm \mu) \mp i\pi\}$. By using 
the identity (\ref{CothIdentity}), the summation over $n$ can be carried out as
\begin{equation}\label{I0Evaluated}
I_0^{\pm} = \sum_{\k} \frac{\beta}{2\omega}
\left[\frac{1}{2} - \frac{1}{e^{\beta(\omega \pm \mu)} + 1} \right ] ~.
\end{equation}
The anti-particle contributions are contained in the term $I_0^{+}$. So by 
ignoring the anti-particle contributions, the divergent zero-point energy and 
by 
using the approximation $\beta\mu \gg 1$, $I_0$ can be evaluated as $I_0 = - 
\beta V \kfb^2/8\pi^2$. Therefore, the ensemble average becomes
\begin{equation}\label{SMinusSquaredAverageFourierSpaceFinal}
\langle {(S_{-}^{\beta})}^2 \rangle = \frac{\beta V e^2\kfb^2}{4\pi^2}
\left(\frac{\kfb^2}{4\pi^2} +  \frac{1}{3\beta^2} \right) ~.
\end{equation}
We note that the Eq. (\ref{SMinusSquaredAverageFourierSpaceFinal}) differs from 
an analogous expression, describing the contributions from the 
electron-electron interaction, given in the textbook by Kapusta and Gale (Eq. 
5.59) \cite{book:kapusta}. However, the expression in the textbook is erroneous 
as it implies that electromagnetic repulsion between the electrons causes a 
reduction of pressure for a system of degenerate electrons in the
ultra-relativistic regime. In particular, if one ignores temperature 
corrections, in the ultra-relativistic limit ($k_F \gg m$) \emph{rhs} of 
the Eq. (\ref{SMinusSquaredAverageFourierSpaceFinal}) varies as
$k_F^4$ whereas the textbook expression varies as $-k_F^4$. On the other hand, 
in non-relativistic limit ($k_F \ll m$), the Eq. 
(\ref{SMinusSquaredAverageFourierSpaceFinal}) varies as
$k_F^6/m^2$ whereas the textbook expression varies as $k_F^4$. We note that the 
textbook expression which describes pressure corrections due to repulsive
electron-electron interaction, changes sign as one goes from relativistic to 
non-relativistic regime. This aspect itself signals internal inconsistency of 
the expression given in the textbook.

\subsubsection{Electron-nuclei interaction}

The leading order contribution due to the electron-nuclei interaction can be 
expressed as
\begin{equation}\label{SPlusAverageRealSpace}
\langle S_{+}^{\beta} \rangle = -Z e^2 d^2 
\int_0^{\beta}d\tau \int d^3\x  J_{\mu}(\tau,\x) \langle
\overline{\psi}(\tau,\x)\gamma^{\mu}\psi(\tau,\x)\rangle ~.
\end{equation}
In order to evaluate the integral (\ref{SPlusAverageRealSpace}), it is 
convenient to express it in the Fourier domain as
\begin{equation}\label{SPlusAverageFourierSpace}
\langle S_{+}^{\beta} \rangle = -Z e^2 d^2  \tilde{J}_{\mu}(\beta) \sum_{n,\k} ~
\mathrm{Tr} \left[\gamma^{\mu}\mathcal{G}(\omega_n,\k) \right] ~,
\end{equation}
where the \emph{average} background 4-current density
\begin{equation}\label{AverageCurrentDensityFourierSpace}
\tilde{J}_{\mu}(\beta) = \frac{1}{\beta V}
\int_0^{\beta}d\tau \int d^3\x ~ J_{\mu}(\tau,\x) ~.
\end{equation}
Within the given box, the spatial motion of the heavier nuclei can be neglected.
So we may assume that the average background 3-current density 
$\tilde{J}^{k}(\beta) = 0$ for $k=1,2,3$. By identifying the average background 
charge density $n_{+} =\tilde{J}^0(\beta) =-\tilde{J}_0(\beta)$ and
by using the trace identity of Dirac matrices, we can express the Eq. 
(\ref{SPlusAverageFourierSpace}) as
\begin{equation}\label{SPlusAverageFourierSpaceEvaluated0}
\langle S_{+}^{\beta} \rangle =  -4 Z e^2 d^2 n_{+} I_2  ~~,~~ 
I_2 = \sum_{n,\k} \frac{p_0}{p^2+m^2}  ~.
\end{equation}
Similar to the Eq. (\ref{I0Evaluated1}), $I_2$ can be expressed as
\begin{equation}\label{I2Evaluated1}
I_2 = \sum_{n,\k} \frac{1}{2} 
\left[\frac{1}{-p_0 + \omega} - \frac{1}{p_0 + \omega}  \right]
\equiv I_2^{-} - I_2^{+}   ~,
\end{equation}
and the summation over the Matsubara frequencies can be carried out as
\begin{equation}\label{I2Evaluated2}
I_2^{\pm} = \sum_{\k} \frac{\beta}{2}
\left[\frac{1}{2} - \frac{1}{e^{\beta(\omega \pm \mu)} + 1} \right ] ~.
\end{equation}
As earlier, by ignoring the anti-particle contributions $I_2^{+}$, the 
divergent zero-point energy and by using the approximation $\beta\mu 
\gg 1$, finite part of $I_2$ can be expressed as $I_2 = - \beta V 
k_F^3/12\pi^2$. The ensemble average $\langle S_{+}^{\beta} \rangle$ 
then becomes
\begin{equation}\label{SPlusAverageFourierSpaceEvaluated}
\langle S_{+}^{\beta} \rangle =  \frac{\beta V Z e^2 d^2  k_F^3
n_{+}}{3\pi^2}  ~.
\end{equation}

\subsubsection{Total contributions from the interactions}

The number density of positively charged nuclei must satisfy $Z n_+  = n_e$
as system is overall electrically neutral. Therefore, by combining 
the contributions from the self-interaction of the electrons
(\ref{SMinusSquaredAverageFourierSpaceFinal})
and the electron-nuclei interaction (\ref{SPlusAverageFourierSpaceEvaluated}), 
we can express the partition function 
(\ref{LogPartitionFunctionInteraction}) due to the total interaction as
\begin{equation}\label{LogPartitionFunctionInteractionTotal}
\ln\mathcal{Z}_{I} = \frac{\beta V e^2 }{96\pi^4}  
\left(3\kfb^4 - 32\pi^2 d^2 n_e k_F^3 \right) ~,
\end{equation}
where we have ignored the finite temperature corrections within the parenthesis 
as the coupling constant $e$ and the term  $(\beta k_F)^{-1}$ both are small.

\subsection{Equation of state}

Using the evaluated partition function we can compute the pressure and the mass 
density within the considered box located at the given radial coordinate. 
Subsequently, we may read off the corresponding equation of state of the 
degenerate matter at the given radial location. For later convenience, we now 
define following dimensionless parameters
\begin{equation}\label{DimensionlessParameter}
 \sigma \equiv \frac{m}{k_F} ~,~
 \sigma_{\mu} \equiv \frac{\mu}{k_F} ~,~
 \sigma_{k} \equiv \frac{\kfb}{k_F}  ~,~
 \sigma_{T} \equiv \frac{k_B T}{k_F}   ~.
\end{equation}
We note that $\sigma^{-1}$, as defined here, can be identified with the so 
called `relativity parameter' in the literature \cite{salpeter1961energy}. 
We also note that $\sigma_{\mu} = \sqrt{1+\sigma^2}$, $\sigma_{k}^2 = 
\sigma_{\mu} - \sigma^2 \arcsinh(1/\sigma)$ and $\sigma_T = (\beta k_F)^{-1}$.
For typical white dwarfs, the Eq. (\ref{BetaKfRatio}) 
implies $\sigma_T \ll 1$. For a system of ultra-relativistic degenerate 
electrons $\sigma \ll 1$ which leads to $\sigma_{\mu} \simeq 1$ and $\sigma_{k} 
\simeq 1$.

\subsection{Number density}

\begin{figure}
\begin{center}
\includegraphics[height=7cm, width=9cm]{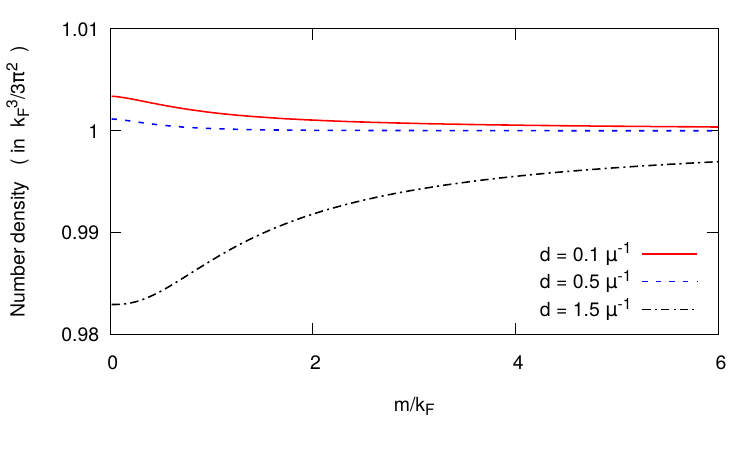}
\end{center}
\caption{The number density of electrons at zero temperature for 
different values of length scale $d$.}
\label{fig:number-density}
\end{figure}

The number density of the electrons can be computed from total partition 
function as $n_e \equiv \langle N\rangle/V = (\beta V)^{-1} (\partial \ln 
\mathcal{Z}/\partial \mu)$.  Given the partition function due to the 
interaction 
terms (\ref{LogPartitionFunctionInteractionTotal}) itself depends on the 
electron number density, it leads to an algebraic equation for $n_e$ as given 
below
\begin{equation}\label{ElectronNumberDensityFull0}
 n_e = \frac{k_F^3}{3\pi^2} \left[ 1 +  \frac{2 +\sigma^2}{(6\sigma_T^2)^{-1}}
 + \frac{3\alpha}{2\pi} \left\{\sigma_k^2 
 -\frac{8\pi^2 d^2 n_e \sigma_{\mu}}
 { k_F }  \right \} \right] ~.
\end{equation}
In order to arrive at the Eq. (\ref{ElectronNumberDensityFull0}), we have 
used two very useful relations $(\partial k_F/\partial \mu) = (\mu/k_F)$ and 
$(\partial \kfb^2/\partial \mu) = 2 k_F$. The Eq.  
(\ref{ElectronNumberDensityFull0}) can be solved in a straightforward manner to 
result
\begin{equation}\label{ElectronNumberDensityFull}
 n_e = \frac{k_F^3}{3\pi^2} \left[ \frac{1 + 6\sigma_T^2 (2 +\sigma^2) 
 + (3\alpha/2\pi) \sigma_k^2 } {1+ d_{+}^2 (\alpha/\pi) \sigma_{\mu}^{-1} } 
\right] ~,
\end{equation}
where the \emph{fine structure constant} is $\alpha = e^2/4\pi$ in natural 
units. By using the chemical potential $\mu$ which comes naturally in the 
partition function (\ref{PartionFunction0}), we have defined a dimensionless
parameter $d_{+} \equiv 2 d \mu$. The parameter $d_{+}$ characterizes the 
associated length scale with the electron-nuclei interaction and it needs to be 
fixed by separate consideration (see Fig. \ref{fig:number-density}).

\subsection{Pressure and energy density}

\begin{figure}
\begin{center}
\includegraphics[height=7cm, width=9cm]{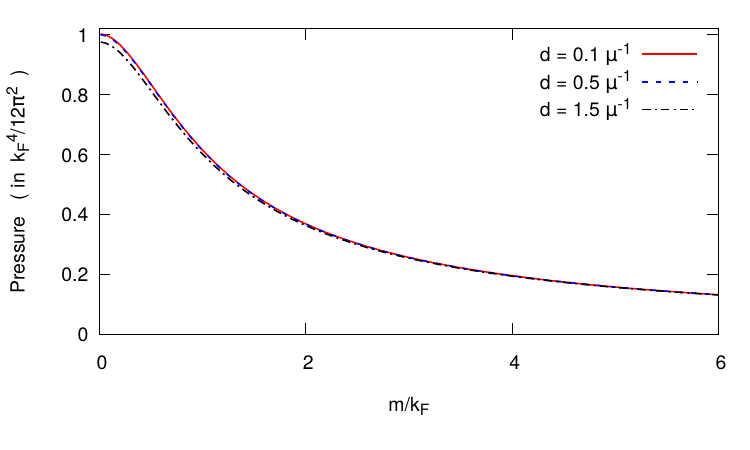}
\end{center}
\caption{The degeneracy pressure at zero temperature for 
different values of length scale $d$.}
\label{fig:pressure}
\end{figure}

Using the expression of pressure for a grand canonical ensemble, we may read 
off the pressure due to the degenerate electrons as $P_{\psi} = (\beta 
V)^{-1}\ln \mathcal{Z}_{\psi}$ and due to the interactions as $P_{I} = (\beta 
V)^{-1}\ln \mathcal{Z}_{I}$. One may check that the radiation pressure 
$P_A = \tfrac{1}{45}\pi^2 \beta^{-4}$ is insignificant even compared to $P_I$, 
as for white dwarfs $\beta k_F \gg 1$. Therefore, by ignoring the radiation 
component, we can express the total pressure which includes leading order 
corrections due to the finite temperature and the fine-structure constant, as
\begin{equation}\label{PressureFull}
P = \frac{k_F^4}{12\pi^2} \left[ \frac{1+24\sigma_T^2}{\sigma_{\mu}^{-1}}
- \frac{3\sigma^2 \sigma_k^{2}}{2} + \frac{3\alpha}{2\pi}
( \sigma_k^4 - \frac{8\pi^2 d_{+}^2 n_e}{3 k_F^3 \sigma_{\mu}^2} ) 
\right]  ~.
\end{equation}
We may again note that the degeneracy pressure depends on the parameter $d_{+}$ 
which characterizes the electron-nuclei interaction length scale (see 
Fig. \ref{fig:pressure}).
The contribution to the total energy density $\rho$ of the system due to 
the degenerate electrons can be computed using the formula $(\rho_e  - \mu 
n_e)V = -(\partial \ln \mathcal{Z} /\partial \beta)$ and which is given by
\begin{equation}\label{ElectronEnergyDensity}
\rho_{e} = - P + \mu n_{e}  ~.
\end{equation}
For a white dwarf, however, the dominant contribution to the total energy 
density comes from the contribution of the nuclei, given by
\begin{equation}\label{MassDensity}
\rho_{n} = (A/Z) m_u n_e  ~,
\end{equation}
where $A$ is the atomic mass number and $m_u$ is the atomic mass unit. For a 
white dwarf with pure Helium ${}^4\mathrm{He}$ core $(A/Z)$ is $2$. Using the 
Eqs. (\ref{ElectronNumberDensityFull}, \ref{PressureFull}, \ref{MassDensity}), 
in principle, one can express the equation of state for the degenerate matter 
within white dwarfs as $P = P(\rho)$ which includes the corrections due to the 
fine structure constant $\alpha$ and the finite temperature.

\subsection{Non-relativistic limit}

We have the considered matter field actions to be manifestly Lorentz 
invariant here. Consequently the studied equation of state is well suited for 
describing the relativistic regime. However, for the consistency, the equation 
of state must also have correct non-relativistic limit when $k_F \ll m$. In 
such limit $\sigma \gg 1$, $\sigma_{\mu} = \sigma + \tfrac{1}{2} \sigma^{-1} - 
\tfrac{1}{8} \sigma^{-3} + \mathcal{O}(\sigma^{-5})$ and $\sigma_{k}^2 = 
\tfrac{2}{3} \sigma^{-1} - \tfrac{1}{5} \sigma^{-3} + 
\mathcal{O}(\sigma^{-5})$. Therefore, in the non-relativistic regime, the 
number density (\ref{ElectronNumberDensityFull}) reduces to
\begin{equation}\label{NRLimitNumberDensity}
n_e \simeq \frac{k_F^3}{3\pi^2} \left[1 + \frac{6 m^2 k_B^2 T^2}{k_F^4} +
\frac{\alpha(1-d_{+}^2) k_F}{\pi m} \right]   ~,
\end{equation}
and the pressure (\ref{PressureFull}) reduces to
\begin{equation}\label{NRLimitPressure}
P \simeq \frac{k_F^5}{15\pi^2 m} \left[1 + \frac{30 m^2 k_B^2 T^2}{k_F^4} 
+ \frac{2\alpha(1- 2d_{+}^2) k_F}{3\pi m}  \right] ~.
\end{equation}
If one disregards the corrections due to the finite temperature and the 
fine-structure constant, the Eqs. (\ref{NRLimitNumberDensity}, 
\ref{NRLimitPressure}) represent the standard non-relativistic expressions. 
However, one may note that in the non-relativistic regime the effects of 
finite temperature become important. Nevertheless, these equations  are valid 
in 
non-relativistic regime as long as corresponding chemical potential $\mu$ 
satisfies $\beta\mu \gg 1$.

\subsection{Temperature corrections}

For non-interacting, zero temperature degenerate electron gas, the number 
density of electrons is given by $n_e = (k_F^3/3\pi^2)$. However, the effect of 
finite temperature causes this relation to be modified even for non-interacting 
electrons as
\begin{equation}\label{ElectronNumberDensityTempCorrection}
\frac{n_e(T)}{n_e(T=0)} = 1 + \frac{6 (2 +\sigma^2) k_B^2 T^2}{k_F^2}  ~.
\end{equation}
Analogously, the effect of finite temperature on the pressure of 
non-interacting degenerate electron gas can be expressed as
\begin{equation}\label{PressureTempCorrection}
\frac{P(T)}{P(T=0)} = 1 + \frac{48 \sigma_{\mu} k_B^2 T^2}
{ k_F^2 (2\sigma_{\mu} - 3\sigma^2 \sigma_k^{2})}  ~.
\end{equation}
Clearly, the finite temperature causes the pressure to increase for a given 
Fermi momentum $k_F$. However, the increase in pressure is very small given it 
is of the order $\sim (\beta k_F)^{-2} \sim 10^{-10}$ for typical white dwarfs 
(\ref{BetaKfRatio}).

\subsection{Fine-structure constant corrections}

The effects of the electromagnetic interaction \emph{i.e.} the Coulomb effects 
on the equation of state are expressed using the fine-structure constant 
$\alpha \simeq 1/137$ which is a small number. However, theses corrections are 
much larger compared to the temperature corrections. At the zero temperature, 
the leading order effect of the fine-structure constant on the electron number 
density can be expressed as
\begin{equation}\label{ElectronNumberDensityAlphaCorrection}
\frac{n_e(\alpha)}{n_e(\alpha=0)} =  1  + \frac{\alpha}{2\pi}  
\left( 3\sigma_k^2 - 2 d_{+}^2 \sigma_{\mu}^{-1} \right) ~.
\end{equation}
Similarly, at the zero temperature the leading order effect of the 
fine-structure constant on the pressure can be expressed as
\begin{equation}\label{PressureAlphaCorrection}
\frac{P(\alpha)}{P(\alpha=0)} = 1 + \frac{\alpha}{3\pi} 
\frac{ (9\sigma_k^4 - 8d_{+}^2 \sigma_{\mu}^{-2})} {(2\sigma_{\mu} - 3\sigma^2 
\sigma_k^{2})}  ~.
\end{equation}
We note that the number density and the pressure both contain an undetermined 
dimensionless parameter $d_{+} = 2d\mu$ in the corrections involving the 
fine-structure constant. As mentioned earlier, the length scale $d$ is 
associated with the current-current interaction between the electrons and the 
nuclei. In the partition function, a natural length scale is provided by the 
chemical potential $\mu$. Therefore, intuitively one would expect that the 
dimensionless parameter $d_{+}$ to be an $\mathcal{O}(1)$ number for the white 
dwarfs. However, determination of its exact numerical values can only be done 
by using separate considerations, possibly by using observations. In the 
standard literature, this one-parameter uncertainty is  often overlooked as 
usually there one fixes the lattice scale associated with positively charged 
nuclei by heuristic arguments. However, we have argued that this length scale is 
associated with the electron-nuclei interaction and its independent 
determination in principle can allow one to understand the property of the 
underlying lattice structure formed by the nuclei within the degenerate matter 
of the white dwarfs.

\subsection{Comparison with Salpeter's corrections}

In order to compare the number density (\ref{ElectronNumberDensityFull}) and 
the pressure (\ref{PressureFull}) with that of Salpeter's we need to set 
$\sigma_T=0$ as these are studied at zero temperature by Salpeter 
\cite{salpeter1961energy}. Further, for comparison we consider the terms up to 
leading order in fine structure constant $\alpha$ from the combined expressions 
of non-interacting degeneracy pressure $P_0$, classical Coulomb corrections 
$P_C$, Thomas-Fermi corrections $P_{TF}$, exchange corrections $P_{ex}$ and 
correlation corrections $P_{cor}$ as described in \cite{salpeter1961energy}.

In the treatment by Salpeter, the relation between the number density of 
electrons $n_e$ and the Fermi momentum $k_F$ is assumed to be fixed. On the 
other hand, the usage of grand canonical partition function here implies that 
there is a modification  to the expression of the electron number density 
due to the electromagnetic interactions. In turns, it would imply a difference 
in equation of state even if the pressure expressions considered by Salpeter 
and here, were to agree.

\subsubsection{No interaction}

The expressions of the number density (\ref{ElectronNumberDensityFull}) and the 
pressure (\ref{PressureFull}) agree exactly with the Salpeter's expressions 
when one ignores the fine-structure constant corrections by setting 
$\alpha=0$ and identifies $\sigma^{-1}$ as the `relativity parameter' $x$.

\subsubsection{Relativistic domain}

\begin{figure}
\begin{center}
\includegraphics[height=7cm, width=9cm]{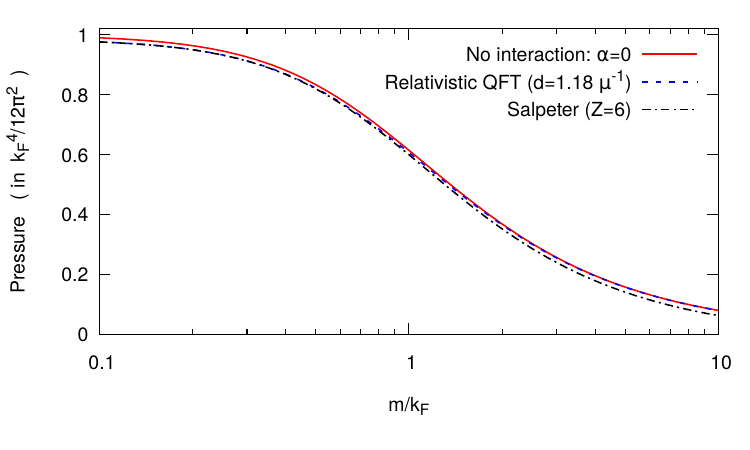}
\end{center}
\caption{A comparison of the pressure in a broadly relativistic domain.}
\label{fig:comparison-ur}
\end{figure}

In the ultra-relativistic limit, $k_F \gg m$, we can express the total pressure 
which includes leading order corrections due to the fine-structure constant, as
\begin{equation}\label{PressureUltraRelativistic}
P = \frac{k_F^4}{12\pi^2} \left[ 1 + \alpha \left(\frac{3}{2\pi}  -
\frac{4 d_{+}^2}{3\pi} \right) \right]  ~.
\end{equation}
The analogous expression for pressure with leading order corrections considered 
by Salpeter, can be expressed as $P = \frac{k_F^4}{12\pi^2} \left[ 1 + \alpha 
(\frac{1}{2\pi} - \frac{6}{5} (\tfrac{4}{9\pi})^{1/3} Z^{2/3}) \right]$ 
\cite{salpeter1961energy}. Therefore, if one chooses $d_{+}^2 = \tfrac{3}{4} + 
(\tfrac{81\pi^2}{250})^{1/3} Z^{2/3}$ then one would get the same pressure 
corrections in the ultra-relativistic limit. In particular, if one chooses the 
atomic number $Z=2$ (Helium) or $Z=6$ (Carbon) then the Salpeter's corrections 
would correspond to the length scale $d$ being $0.88 ~\mu^{-1}$, $1.18 
~\mu^{-1}$ respectively. This is in agreement with the intuitive expectation 
that $d_{+}$ should be an $\mathcal{O}(1)$ number. The pressure comparison 
in a broadly relativistic domain is given in the Fig. \ref{fig:comparison-ur}.

Nevertheless, we emphasize that the length scale $d$ is undetermined apriori in 
the approach that we have considered here. For a given system of degenerate 
electrons and nuclei, in principle, it may be possible to derive an 
\emph{effective action} corresponding to Eq. (\ref{InteractionActionNucleon}) 
where $\sim d_{+} e$ may be viewed as the renormalized coupling constant between 
the electrons and the nuclei at the energy scale set by the chemical potential 
$\mu$.

\subsubsection{Non-relativistic domain}

\begin{figure}
\begin{center}
\includegraphics[height=7cm, width=9cm]{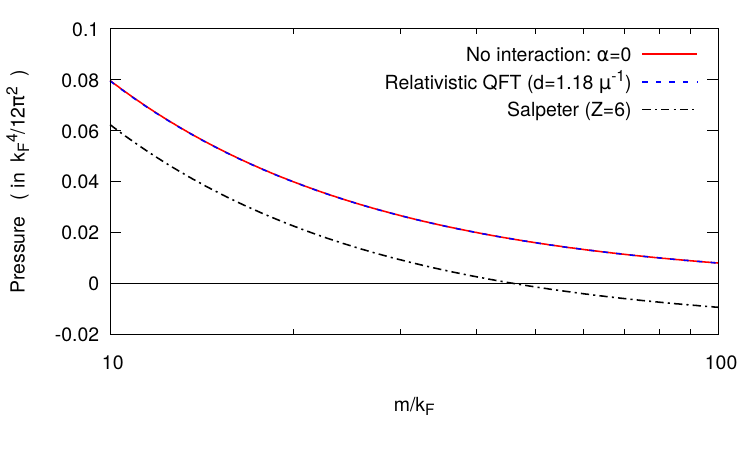}
\end{center}
\caption{A comparison of the pressure in a broadly non-relativistic domain. The 
pressure becoming negative signals the breakdown of the underlying assumptions 
in Salpeter's treatment.}
\label{fig:comparison-nr}
\end{figure}

The corrections to the pressure expression, considered by Salpeter, are known 
to become unreliable in a fairly non-relativistic domain. Salpeter noted that 
with such corrections the total pressure could become negative, signaling the 
breakdown of the underlying assumptions \cite{salpeter1961energy}. In contrast, 
the non-relativistic expression (\ref{NRLimitPressure}) here remains well 
defined. A comparison of the pressure in a broadly non-relativistic domain is 
given in the Fig. \ref{fig:comparison-nr}.

\section{Degenerate plasma in curved spacetime}\label{DegenerateCurved}

So far we have studied the degenerate plasma by considering the Minkowski 
spacetime which has vanishing spacetime curvature. In such a set up, the 
gravitational interaction is incorporated by using the Newtonian gravity. 
For example, the celebrated Chandrasekhar mass limit for the white dwarfs is 
derived using the combination of the Newtonian gravity and the equation of state 
of the degenerate plasma computed in the Minkowski spacetime.

However, the framework of Newtonian gravity is not a very accurate description 
when the strength of gravitational interaction is strong. On the other hand, 
Einstein's general relativity is known to be the most accurate classical 
description of the gravitational interaction where the gravitation manifests 
itself through the curvature of the spacetime. Among the compact astrophysical 
objects, the effect of general relativity for the white dwarfs is relatively 
smaller. However, the general relativistic effect on a neutron star is quite 
significant. On the mass-radius relation of these compacts stars, the effects 
of general relativity appear in two different ways. The first effect arises 
from the usage of Einstein's field equations rather than using Poisson's 
equation of the Newtonian gravity. The second effect arises due to the effect of 
curved spacetime on the matter field dynamics and the resultant modifications 
to the matter EOS due to the gravitational time dilation.

\subsection{Tolman-Oppenheimer-Volkoff Equations}

In order to include the effects of general relativity in the study of these 
compact stars, for simplicity, we assume that the spacetime of these stars can 
be described by a spherically symmetric spacetime geometry. This assumption 
then leads the exterior vacuum spacetime around these stars to be described by 
the Schwarzschild solution of Einstein's equation. On the other hand, in the 
interior spherical geometry, the invariant line element can be expressed
using the natural units, $c=\hbar=1$, as
\begin{equation}\label{InteriorMetric}
ds^{2} = -e^{2\Phi(r)}dt^{2} + e^{2\nu(r)}dr^{2} 
+ r^{2} (d\theta^{2} + \sin^{2}\theta d\phi^{2}) ~,
\end{equation}  
where $\Phi(r)$ and $\nu(r)$ are the metric functions that are governed by 
Einstein's field equations. Further, we assume that the matter stress-energy 
tensor inside these stars can be treated as a perfect fluid whose expression is 
given by
\begin{equation}\label{PerfectFluid}
T^{\mu\nu} = (\rho+P)u^{\mu}u^{\nu} + Pg^{\mu\nu}  ~,
\end{equation}
where $P$ is the pressure, $\rho$ is the energy density, $g_{\mu\nu}$ 
is the spacetime metric (\ref{InteriorMetric}), and $u^{\mu}$ is the 4-velocity 
of the co-moving observer. The Einstein equations corresponding to the interior 
metric (\ref{InteriorMetric}) together with covariant conservation equations of 
the stress-energy tensor (\ref{PerfectFluid}) lead to the following 
set of equations
\begin{equation}\label{TOVEq}
\frac{d\Phi}{dr}=\frac{G(\mathcal{M} + 4\pi r^{3}P)}{r(r - 2G\mathcal{M})} 
~,~ \frac{dP}{dr} = - (\rho + P)\frac{d\Phi}{dr} ~,
\end{equation}
where $\mathcal{M} \equiv \mathcal{M}(r)$, often referred to as the `enclosed 
mass' upto the radial coordinate $r$, satisfies the equation
\begin{equation}\label{MEq}
\frac{d\mathcal{M}}{dr}  = 4 \pi r^2 \rho  ~.
\end{equation}
These equations (\ref{TOVEq}, \ref{MEq}) govern the matter distribution inside
the spherical stars and are known as the Tolman-Oppenheimer-Volkoff (TOV) 
equations. In the limit where the speed of light $c\rightarrow\infty$, the TOV 
eqs. lead to the Poisson equation of the Newtonian gravity. The other metric 
function $\nu(r)$ can be partially solved and can be expressed as
\begin{equation}
e^{-2\nu(r)} = 1-\frac{2 G \mathcal{M}}{r} ~.
\end{equation}
In order to solve the TOV equations where there are three independent equations 
(\ref{TOVEq}, \ref{MEq}) but four unknown functions, namely $\mathcal{M}$, 
$\Phi$, $P$ and $\rho$, one needs to provide the matter equation of state of the 
form $P = P(\rho)$. However, being nonlinear in nature, these equations cannot 
be solved analytically for a given matter EOS. Therefore, in order to compute 
the resultant mass-radius relations of these compact stars, one needs to deploy 
numerical methods for solving the TOV equations.

\subsection{Effect of gravitational time dilation on EOS}

The second kind of general relativistic effect that arises in the study of 
compact astrophysical objects, is the effect of gravitational time dilation on 
the matter EOS. In the literature, this effect is often overlooked. In the 
general relativity, any curved spacetime can be described locally by a flat 
metric. This argument is then used to deploy a matter EOS computed in a 
\emph{globally} flat spacetime, here referred to as the \emph{flat} EOS, for 
solving the TOV equations. However, such a argument misses the fact that inside 
the star, the metric function $\Phi$ varies radially over the scale of the star, 
as governed by the eq. (\ref{TOVEq}). Therefore, two locally inertial frames 
which are located at different radial locations have different lapse functions 
and consequently have different clock speeds.

By using a first-principle approach, the effect of gravitational time dilation 
on the matter EOS has been computed for different types of degenerate matter 
\cite{hossain2021equation, hossain2021higher}. Here we follow a simpler 
approach by suitably transforming the quantities computed in a globally flat 
spacetime, as prescribed in the ref. \cite{hossain2021equation}. We would like 
to emphasize here that in this context two different scales are involved. 
Firstly, at the scale of local thermodynamical equilibrium, the thermodynamical 
quantities such as the pressure and the energy density are uniform. In this 
scale, the metric function $\Phi$ must also be considered as uniform. 
For definiteness, let us consider a small box whose center is located at a 
radial coordinate $r_0$, so that local thermodynamical equilibrium holds 
within the box. For a spherical star, one can always define a local set of 
coordinates as
$\tilde{X} = e^{\nu(r_0)} r \sin \bar{\theta} \cos\phi$, 
$\tilde{Y} = e^{\nu(r_0)} r \sin \bar{\theta} \sin\phi$, and
$\tilde{Z} = e^{\nu(r_0)} r \cos \bar{\theta}$ where $\bar{\theta} =  
e^{-\nu(r_0)}\theta$ such that the metric (\ref{InteriorMetric}) inside the box 
can be written as \cite{hossain2021equation}
\begin{equation}\label{MetricInTOVBox}
ds^2 = - e^{2\Phi(r_0)}dt^2 + d\tilde{X}^2 + d\tilde{Y}^2 + d\tilde{Z}^2 ~.
\end{equation}
One may now use the locally flat coordinate system 
$(t,\tilde{X},\tilde{Y},\tilde{Z})$ within the box 
to compute the matter EOS within the box where the metric function $\Phi \equiv 
\Phi(r_0)$ has a fixed value. This computation leads the matter EOS to retain 
its dependence on the metric function $\Phi$. Subsequently, such an EOS, 
referred to as the \emph{curved} EOS, can be used to solve the TOV equations. 
Nevertheless, it is possible to obtain the corresponding curved EOS starting 
from the partition function computed in a globally flat spacetime. We note that 
if one defines a new time coordinate as $\tilde{t} = e^{\Phi}t$ within the box 
then the metric in the coordinate system 
$(\tilde{t},\tilde{X},\tilde{Y},\tilde{Z})$ becomes the same as the Minkowski 
metric. At thermodynamical equilibrium, the anti-periodic boundary condition 
(\ref{FermionicBoundaryCondition}) then leads to the following relations 
\cite{hossain2021equation}
\begin{equation}\label{BetaMuRelations}
\tilde{\beta} = \beta e^{\Phi}  ~~,~~ \tilde{\mu} =  \mu e^{-\Phi} ~,
\end{equation}
where $\tilde{\beta} = 1/(k_B \tilde{T})$ with $\tilde{T}$ and $\tilde{\mu}$ 
being the temperature and chemical potential respectively, as seen in the 
frame $(\tilde{t},\tilde{X},\tilde{Y},\tilde{Z})$. Using the transformations 
(\ref{BetaMuRelations}) and by ignoring the contributions coming from the 
electromagnetic interactions, the partition function for degenerate electrons 
(\ref{LogPartitionFunctionFermionFinal}) within the box becomes
\begin{equation}\label{LogPartitionFunctionCurved}
\ln\mathcal{Z}_{\psi} = \frac{\beta V e^{-3\Phi}}{24\pi^2} \left[ 
2\mu \mu_{m}^3 - 3\mt^2 \muimb^2 
+ \frac{48 \mu \mu_{m}}{\beta^{2}} \right] ~,
\end{equation}
where $\mu_{m} \equiv \sqrt{\mu^2 - \mt^2}$, $\muimb^2 \equiv \mu\mu_{m} - \mt^2 
\ln(\tfrac{\mu + \mu_{m}}{\mt})$ and $\mt =  m e^{\Phi}$. Here we have expressed 
the partition function in terms of the chemical potential $\mu$, as $\mu$ is independent of the choice of spatial coordinates, unlike the Fermi 
momentum $k_F$.

By using the expression of the partition function 
(\ref{LogPartitionFunctionCurved}) for degenerate electrons, the expressions for 
the number density $n_{e}$, the pressure $P$ and the energy density $\rho_e$ can 
be derived explicitly as $n_e = (\beta V) ^{-1} (\partial \ln \mathcal{Z} 
_{\psi} /\partial \mu)$, $P = (\beta V)^{-1}\ln \mathcal{Z}_{\psi}$, and 
$(\rho_e - \mu n_e)V = -(\partial \ln \mathcal{Z}_{\psi} /\partial \beta)$. 
The expressions for the pressure and the energy density then can be written as
\begin{align}\label{PressureCurved}
P & = e^{\Phi}\frac{m^4}{24\pi^2} \Big[ \sqrt{(b n_e)^{2/3} + 1}
\left\{2(b n_e) - 3(b n_e)^{1/3} \right\} \nonumber\\
 & + 3 \arcsinh( (b n_e)^{1/3} )\Big] ~,~ 
\end{align}
and
\begin{equation}\label{EnergyDensityCurved}
\rho_e = -P + e^{\Phi} \frac{m^4}{3\pi^2} \sqrt{(b n_e)^{2/3} + 1}
~ (b n_e) ~,
\end{equation}
where $b = (3\pi^2/m^3)(Z/A)$. On the other hand, in the curved spacetime the 
expression of the energy density of the nuclei becomes 
\begin{equation}\label{MassDensityCurved}
\rho_{n} = (A/Z) m_u n_e e^{\Phi} ~,
\end{equation}
when one includes the effect of gravitational time dilation 
\cite{hossain2021equation}. We note that in the eqs. (\ref{PressureCurved}, 
\ref{EnergyDensityCurved}, \ref{MassDensityCurved}) if one turns off the 
gravitational time dilation effect by setting $e^{\Phi}\to 1$, then the 
expressions of the pressure and the energy density derived in curved spacetime 
reduce to their Minkowski spacetime counterparts (\ref{PressureFull}, 
\ref{ElectronEnergyDensity}, \ref{MassDensity}) without the corrections from 
the electromagnetic interactions.

\section{White dwarfs}

In this section, we discuss a few consequences of using the equation of state
of the degenerate plasma which is computed using the thermal field theory 
approach, for the white dwarfs. The expression for the degeneracy 
pressure which is relevant for the white dwarfs, is obtained in eq. 
(\ref{PressureFull}) for the flat spacetime and in the eq. 
(\ref{PressureCurved}) for the curved spacetime but without the fine-structure 
constant corrections. On the other hand, the expressions for the relevant 
mass-energy density, obtained by considering the appropriate number of 
nuclei present within the white dwarfs to ensure overall $U(1)$ charge 
neutrality, are given in the eqs. (\ref{MassDensity}, \ref{MassDensityCurved}).

\subsection{Modified mass limit of white dwarfs}

The usage of the Newtonian gravity together with the equation of state computed 
in the Minkowski spacetime leads to a maximum mass of a white dwarf, known as 
the Chandrasekhar mass limit. Without the fine-structure constant corrections, 
this mass limit can be expressed as $M^0_{ch} = (4/\sqrt{\pi}) 
(\mathrm{K}_0/G)^{3/2} |\xi_0^2 ~ \theta'(\xi_0)|$ where $K_0$ is known as the 
polytropic constant, given by
\begin{equation}\label{EoSConstant}
\mathrm{K}_0 = \frac{(3\pi^2)^{1/3}}{4 ((A/Z) m_u)^{4/3}} ~.
\end{equation}
Here $\theta$ is the solution of the Lane-Emden equation $\xi^{-2} 
\tfrac{d}{d\xi} \left(\xi^2 \frac{d\theta}{d\xi} \right) = - \theta^3$ 
and $\xi_0$ is the point where $\theta(\xi_0)$ vanishes first. By solving the 
Lane-Emden equation numerically, one obtains $|\xi_0^2 ~\theta'(\xi_0)| \simeq 
2.02$. 

In order to obtain the effect of the fine-structure constant on the 
Chandrasekhar mass limit, we need to find the corrections to the equation of 
state in the ultra-relativistic limit. In such limit, the eqs. 
(\ref{ElectronNumberDensityFull}, \ref{PressureFull}, \ref{MassDensity}) 
together lead to a polytropic equation of state of the form $P \propto 
\rho^{4/3}$ where fine-structure constant modifies the proportionality constant. 
Such a modification in turn leads to the modified Chandrasekhar mass limit 
$M_{ch}$ which including up to the leading order correction in the 
fine-structure constant $\alpha$, is given by \cite{Hossain:2019cml}
\begin{equation}\label{ChandrasekharMassLimit}
\frac{M_{ch}}{M^0_{ch}} = 1 - \frac{3\alpha}{4\pi}  ~.
\end{equation}
We note that the effect of the fine-structure constant $\alpha$ reduces the 
Chandrasekhar mass limit for white dwarfs by a universal factor 
\cite{Hossain:2019cml}. In particular, the length scale $d$ which is associated 
with the electron-nuclei interaction, does not affect the mass limit for the 
white dwarfs. In contrast, the modified Chandrasekhar mass limit which uses 
the Salpeter's corrections, can be expressed up to leading order in $\alpha$, 
as $M_{ch}/M^0_{ch} = 1 - \frac{3}{2} \alpha \left[\frac{6}{5}(\tfrac{4}{9\pi})
^{1/3} Z^{2/3} - \tfrac{1}{2\pi} \right]$
\cite{nauenberg1972analytic,salpeter1961energy,hamada1961models}. So there the
reduction of the Chandrasekhar mass limit is non-universal in nature as it 
depends on the atomic number of the constituent nuclei.

\subsection{Mass radius relations}

\begin{figure}
\begin{center}
\includegraphics[height=7cm, width=9cm]{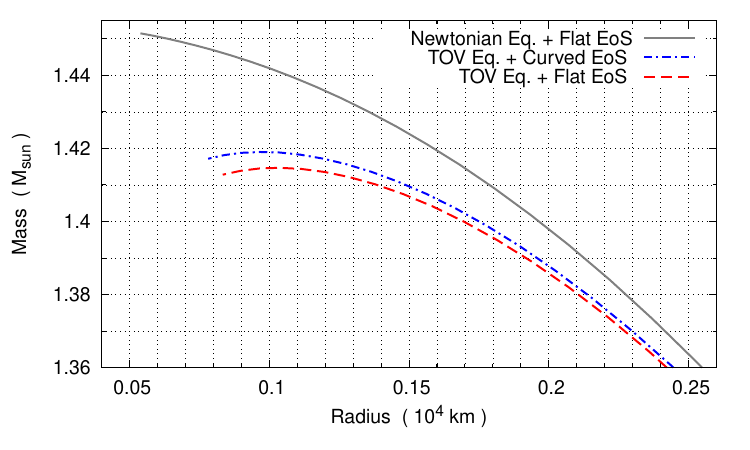}
\end{center}
\caption{Comparison of mass radius relations of the white dwarfs in the 
Newtonian gravity and the general relativity without the corrections involving 
the fine-structure constant.}
\label{fig:comparison-mass-radius-WD}
\end{figure}

In order to obtain the mass-radius relations of the white dwarfs that includes 
modifications from the general relativity, the TOV eqs. (\ref{TOVEq}, 
\ref{MEq}) needs to be solved numerically. For the white dwarfs, 
the dominant contribution to energy density comes from the nuclei. In a 
computation based on a globally flat spacetime, the energy density due to the 
nuclei is given by the eq. (\ref{MassDensity}). On the other hand, in a curved 
spacetime the expression of the energy density due to the nuclei is given in 
the eq. (\ref{MassDensityCurved}) which includes the effect of gravitational 
time dilation. By ignoring the corrections due to the fine-structure constant 
but including two different effects from the general relativity, the resultant 
mass-radius relations for the white dwarfs are plotted in the Fig. 
\ref{fig:comparison-mass-radius-WD}. We note that the maximum mass of the white 
dwarfs reduces by around $2.4\%$ due to the usage of the TOV eqs. compared to 
the same of the Newtonian gravity. On the other hand, the inclusion of the time 
dilation effect into the matter EOS leads to an increase of the maximum mass by 
around $0.7\%$.

\section{Neutron stars}

Unlike the case of white dwarfs, the effect of general relativity on the
neutron stars is significant. It is essentially due to the fact that the 
neutron stars are among the most compact astrophysical objects that can be
observed directly. Such a compact yet massive object curves the spacetime 
around it very strongly. Besides, it requires its interior energy density 
to be order of the nuclear density. Unfortunately, the physical understanding of 
such a dense nuclear matter plasma is still an active area of research. 
Consequently, several models of the equation of states for the neutron stars 
have been studied in the literature.

\subsection{Mass radius relations}

\begin{figure}
\begin{center}
\includegraphics[height=7cm, width=9cm]{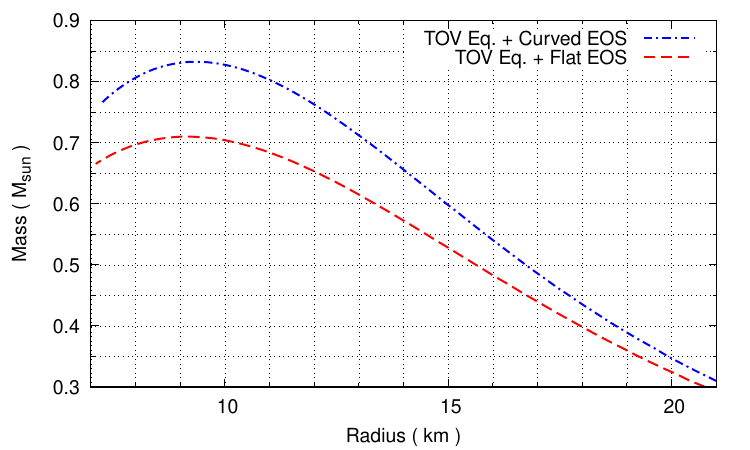}
\end{center}
\caption{Mass radius relations of the ideal neutron stars using both the curved 
EOS and the flat EOS.}
\label{fig:comparison-mass-radius-NS}
\end{figure}

The simplest model of the plasma contained  within the neutron stars 
consists of an ensemble of non-interacting degenerate neutrons. The relevant 
equation of state can be obtained from the eqs. (\ref{PressureCurved}, 
\ref{EnergyDensityCurved}) by mapping the mass of electrons to the mass of 
neutrons and by setting $(Z/A)=1$. The corresponding TOV eqs. then can be 
solved using the numerical methods. The resultant mass-radius relations for 
such an ideal neutron star are plotted in the Fig. 
\ref{fig:comparison-mass-radius-NS} for both the curved EOS and the flat EOS 
\cite{hossain2021equation}. The curved EOS leads to a significant increase in 
the mass of neutron stars for a given radius. In particular, the maximum mass 
of the ideal neutron stars increases by around $16.9\%$ whereas its 
corresponding radius increases by around $2.2\%$.

\begin{figure}
\begin{center}
\includegraphics[height=7cm, width=9cm]{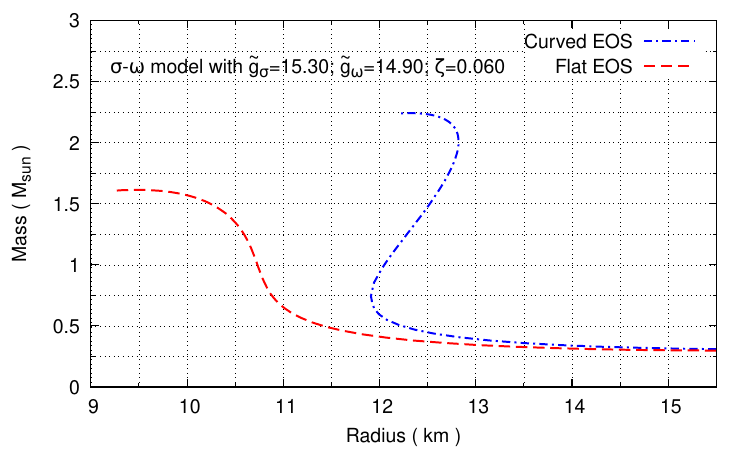}
\end{center}
\caption{Mass radius relations of the neutron stars in the $\sigma-\omega$ 
model using both the curved EOS and flat EOS for a given set of parameter 
values.}
\label{fig:mass-radius-NS}
\end{figure}

Nevertheless, it is quite clear from the Fig. 
\ref{fig:comparison-mass-radius-NS}, that the simplest model of equation of 
state for the neutron stars can not explain the observations of 
astrophysical neutron stars having mass even more than $2M_{\odot}$ 
\cite{linares2018peering, cromartie2020relativistic}. In order to explain such  
high masses of the neutron stars, one often consider the models of equation of 
state of neutron stars where the nuclear plasmas are made of interacting  
nucleons. In the framework of quantum hadrodynamics \cite{serot1992quantum, 
serot1997recent}, one such model to describe the nuclear matter within the 
neutron stars is known as the so-called $\sigma-\omega$ model 
\cite{whittenbury2014quark, katayama2012equation, miyatsu2013new}. By 
considering a  $\sigma-\omega$ model having two \emph{baryons}, namely the 
neutron and the proton, a \emph{lepton} namely the electron, a massive 
\emph{scalar} meson $\sigma$ and a self-interacting \emph{vector} meson 
$\omega$, both the flat EOS and the curved EOS for the neutron stars have been 
computed in \cite{hossain2021higher}. The corresponding mass-radius relations of 
the neutron stars are shown in the Fig. \ref{fig:mass-radius-NS}. The effect of 
gravitational time dilation on EOS leads to a significant increase in the 
maximum mass of neutron stars. In particular, here the maximum mass of the 
neutron stars increases almost by $39.1\%$, and its corresponding radius 
increases by almost $29.8\%$ for a given set of parameters of the 
$\sigma-\omega$ model. 

\section{Discussions}

In summary, in this review, we have employed the methods of relativistic thermal 
quantum field theory to compute the equation of states of degenerate plasmas 
contained within the compact astrophysical objects such as the white dwarfs and 
the neutron stars. In particular, we have computed these equation of states 
including leading order corrections due to the finite temperature, due to the 
electromagnetic interactions, parameterized by the fine-structure constant, 
as well as the effect of gravitational time dilation.
In the literature, the effects of the electromagnetic interaction on the 
equation of state of the degenerate matter within the white dwarfs are computed 
by considering the so-called classical Coulomb energy, the Thomas-Fermi effect, 
the exchange and correlation energy at zero temperature. These computations 
rely on the electrostatic considerations which are non-relativistic ab 
initio. After reviewing the existing treatments in the literature, here we have 
presented the computation of the equation of states of degenerate matter for 
the white dwarfs by employing the methods of thermal quantum field theory 
which is a relativistic approach ab-initio. The resultant correction to the 
equation of state due to the fine-structure constant has two components. The 
first component arises from the self-interaction between the degenerate 
electrons and described by the action of quantum electrodynamics. For the second 
component we have considered a Lorentz invariant interaction term to describe 
the interaction between the electrons and the positively charged nuclei. 
Further, we have argued that a fully relativistic consideration leads to an 
apriori undetermined length scale in the corrections to the equation of state 
involving the electron-nuclei interaction. This aspect of the equation state is 
overlooked in the literature. Instead there one fixes the associated scale by 
using heuristic arguments. An independent determination of this length scale may 
shed light on the underlying lattice structure formed by the nuclei within the 
degenerate matter of the white dwarfs. Besides, the effect of fine-structure 
constant reduces the Chandrasekhar mass limit of the white dwarfs by a universal 
factor which is independent of the atomic number of the constituent nuclei and 
the electron-nuclei interaction length scale.

In order to include the effects of general relativity in the equation of states 
of the degenerate plasmas, here we have considered the spacetime of these 
compact stars to be described by a spherically symmetric spacetime geometry. We 
have shown that the effects of general relativity on the mass radius relation 
of the compacts stars appear in two different ways. The first kind of effect 
arises from the usage of Einstein's field equations rather than using the 
Poisson's equation of Newtonian gravity. The second kind of effect arises due 
to the effect of gravitational time dilation on matter field dynamics and the 
resultant modifications to the matter equation of state.  Among the compact 
astrophysical objects, the effect of general relativity is relatively smaller
for the white dwarfs. However, the general relativistic effect on the neutron 
stars is quite significant.

In this review, we have primarily considered the system of plasmas 
within the astrophysical compacts stars. However, in principle, the field 
theoretical description of the plasmas that are studied here, can also be 
applied for a earth-based plasma system as long as the degeneracy condition 
$\beta \mu \gg 1$ can be satisfied. For examples, if one is dealing with the 
plasmas having temperature $10^{4}$ K and if the associated chemical potential 
$\mu$ is $100~eV$ then $\beta\mu > 10^2$ where the description as studied 
here would be equally applicable.

Finally, we note that the observations of gravitational waves originating from 
coalescence events involving neutron stars, by the current ground-based 
gravitational wave detectors have already started putting severe constraints on 
the equation of states of the degenerate plasmas present within the neutron 
stars \cite{zhang2019extracting, lackey2006observational, 
takami2014constraining, chatziioannou2020neutron}. Furthermore, future detection 
of low-frequency gravitational waves from the extreme mass-ratio merger of a 
black hole and a white dwarf could determine the equation of state of the 
degenerate matter within the white dwarf with an accuracy reaching up to $0.1\%$ 
\cite{Han:2017kre}. It is based on the expectation that the tidal disruption of 
a white dwarf, during the final phase of inspiral around a massive black hole, 
could be measured very accurately through low-frequency gravitational wave 
signals. The properties of the tidal disruption of a white dwarf would 
necessarily depend on the equation of state of the degenerate matter present 
within the white dwarf. On the other hand, we may note from the Eqs. 
(\ref{ElectronNumberDensityAlphaCorrection}, \ref{PressureAlphaCorrection}) that 
the corrections to the number density and the pressure due to the fine-structure 
constant are of the order $\sim \alpha/\pi \sim 0.2\%$. Therefore, the effects 
of the fine-structure constant as studied here could well be within the 
detection realm of the future gravitational wave detectors. We may mention here 
that the deviation of equation of state compared to that of Salpeter's equation 
of state for white dwarf is relatively larger in the non-relativistic regime, as 
can be seen in the Fig. \ref{fig:comparison-nr}. Clearly, such deviations can be 
confronted with the expected accuracy of the future low-frequency gravitational 
wave detectors.

\section{Acknowledgments}

SM would like to thank IISER Kolkata for supporting this work through a 
doctoral fellowship. 

\bibliographystyle{apsrev}

\begin{thebibliography}{26}
\expandafter\ifx\csname natexlab\endcsname\relax\def\natexlab#1{#1}\fi
\expandafter\ifx\csname bibnamefont\endcsname\relax
  \def\bibnamefont#1{#1}\fi
\expandafter\ifx\csname bibfnamefont\endcsname\relax
  \def\bibfnamefont#1{#1}\fi
\expandafter\ifx\csname citenamefont\endcsname\relax
  \def\citenamefont#1{#1}\fi
\expandafter\ifx\csname url\endcsname\relax
  \def\url#1{\texttt{#1}}\fi
\expandafter\ifx\csname urlprefix\endcsname\relax\def\urlprefix{URL }\fi
\providecommand{\bibinfo}[2]{#2}
\providecommand{\eprint}[2][]{\url{#2}}

\bibitem[{Abbott et~al.(2016)Abbott, Abbott, Abbott, Abernathy, Acernese,
  Ackley, Adams, Adams, Addesso, Adhikari et~al.}]{abbott2016observation}
Abbott BP, Abbott R, Abbott T, Abernathy M, Acernese F, Ackley K, Adams C,
  Adams T, Addesso P, Adhikari R, et~al. (2016) Observation of gravitational
  waves from a binary black hole merger. Physical review letters 116(6):061102

\bibitem[{Abbott et~al.(2017)Abbott, Abbott, Abbott, Acernese, Ackley, Adams,
  Adams, Addesso, Adhikari, Adya et~al.}]{abbott2017gw170817}
Abbott BP, Abbott R, Abbott T, Acernese F, Ackley K, Adams C, Adams T, Addesso
  P, Adhikari R, Adya V, et~al. (2017) Gw170817: observation of gravitational
  waves from a binary neutron star inspiral. Physical review letters
  119(16):161101

\bibitem[{Shaddock(2008)}]{shaddock2008space}
Shaddock D (2008) Space-based gravitational wave detection with lisa. Classical
  and Quantum Gravity 25(11):114012

\bibitem[{Hulse and Taylor(1975)}]{hulse1975discovery}
Hulse RA, Taylor JH (1975) Discovery of a pulsar in a binary system. The
  Astrophysical Journal 195:L51--L53

\bibitem[{Taylor and Weisberg(1989)}]{taylor1989further}
Taylor JH, Weisberg JM (1989) Further experimental tests of relativistic
  gravity using the binary pulsar psr 1913+ 16. The Astrophysical Journal
  345:434--450

\bibitem[{\citenamefont{Chandrasekhar}(1931)}]{chandrasekhar1931maximum}
\bibinfo{author}{\bibfnamefont{S.}~\bibnamefont{Chandrasekhar}},
  \bibinfo{journal}{The Astrophysical Journal} \textbf{\bibinfo{volume}{74}},
  \bibinfo{pages}{81} (\bibinfo{year}{1931}).

\bibitem[{\citenamefont{Chandrasekhar}(1935)}]{chandrasekhar1935highly}
\bibinfo{author}{\bibfnamefont{S.}~\bibnamefont{Chandrasekhar}},
  \bibinfo{journal}{Monthly Notices of the Royal Astronomical Society}
  \textbf{\bibinfo{volume}{95}}, \bibinfo{pages}{207} (\bibinfo{year}{1935}).

\bibitem[{\citenamefont{Frenkel}(1928)}]{frenkel1928}
\bibinfo{author}{\bibfnamefont{J.}~\bibnamefont{Frenkel}},
  \bibinfo{journal}{Zeitschrift fur Physik} \textbf{\bibinfo{volume}{50}}, 
pp. \bibinfo{pages}{234--248} (\bibinfo{year}{1928}).

  
\bibitem[{\citenamefont{Kothari}(1938)}]{kothari1938theory}
\bibinfo{author}{\bibfnamefont{D.}~\bibnamefont{Kothari}},
  \bibinfo{journal}{Proceedings of the Royal Society of London. Series A,
  Mathematical and Physical Sciences} pp. \bibinfo{pages}{486--500}
  (\bibinfo{year}{1938}).

\bibitem[{\citenamefont{Auluck and Mathur}(1959)}]{auluck1959electrostatic}
\bibinfo{author}{\bibfnamefont{F.}~\bibnamefont{Auluck}} \bibnamefont{and}
  \bibinfo{author}{\bibfnamefont{V.}~\bibnamefont{Mathur}},
  \bibinfo{journal}{Zeitschrift fur Astrophysik} \textbf{\bibinfo{volume}{48}},
  \bibinfo{pages}{28} (\bibinfo{year}{1959}).

\bibitem[{\citenamefont{Salpeter}(1961)}]{salpeter1961energy}
\bibinfo{author}{\bibfnamefont{E.~E.} \bibnamefont{Salpeter}},
  \bibinfo{journal}{The Astrophysical Journal} \textbf{\bibinfo{volume}{134}},
  \bibinfo{pages}{669} (\bibinfo{year}{1961}).

\bibitem[{\citenamefont{Hamada and Salpeter}(1961)}]{hamada1961models}
\bibinfo{author}{\bibfnamefont{T.}~\bibnamefont{Hamada}} \bibnamefont{and}
  \bibinfo{author}{\bibfnamefont{E.}~\bibnamefont{Salpeter}},
  \bibinfo{journal}{The Astrophysical Journal} \textbf{\bibinfo{volume}{134}},
  \bibinfo{pages}{683} (\bibinfo{year}{1961}).

\bibitem[{\citenamefont{Nauenberg}(1972)}]{nauenberg1972analytic}
\bibinfo{author}{\bibfnamefont{M.}~\bibnamefont{Nauenberg}},
  \bibinfo{journal}{The Astrophysical Journal} \textbf{\bibinfo{volume}{175}},
  \bibinfo{pages}{417} (\bibinfo{year}{1972}).

\bibitem[{\citenamefont{Koester and Chanmugam}(1990)}]{koester1990physics}
\bibinfo{author}{\bibfnamefont{D.}~\bibnamefont{Koester}} \bibnamefont{and}
  \bibinfo{author}{\bibfnamefont{G.}~\bibnamefont{Chanmugam}},
  \bibinfo{journal}{Reports on Progress in Physics}
  \textbf{\bibinfo{volume}{53}}, \bibinfo{pages}{837} (\bibinfo{year}{1990}).

\bibitem[{\citenamefont{Stuart L.~Shapiro}(1983)}]{book:Shapiro.Teukolsky}
\bibinfo{author}{\bibfnamefont{S.~A.~T.} \bibnamefont{Stuart L.~Shapiro}},
  \emph{\bibinfo{title}{Black holes, white dwarfs, and neutron stars: the
  physics of compact objects}} (\bibinfo{publisher}{Wiley},
  \bibinfo{year}{1983}), \bibinfo{edition}{1st} ed.

\bibitem[{\citenamefont{Rotondo
  et~al.}(2011{\natexlab{a}})\citenamefont{Rotondo, Rueda, Ruffini, and
  Xue}}]{rotondo:PhysRevC.83.045805}
\bibinfo{author}{\bibfnamefont{M.}~\bibnamefont{Rotondo}},
  \bibinfo{author}{\bibfnamefont{J.~A.} \bibnamefont{Rueda}},
  \bibinfo{author}{\bibfnamefont{R.}~\bibnamefont{Ruffini}}, \bibnamefont{and}
  \bibinfo{author}{\bibfnamefont{S.-S.} \bibnamefont{Xue}},
  \bibinfo{journal}{Phys. Rev. C} \textbf{\bibinfo{volume}{83}},
  \bibinfo{pages}{045805} (\bibinfo{year}{2011}{\natexlab{a}}).

\bibitem[{\citenamefont{Rotondo
  et~al.}(2011{\natexlab{b}})\citenamefont{Rotondo, Rueda, Ruffini, and
  Xue}}]{rotondo:PhysRevD.84.084007}
\bibinfo{author}{\bibfnamefont{M.}~\bibnamefont{Rotondo}},
  \bibinfo{author}{\bibfnamefont{J.~A.} \bibnamefont{Rueda}},
  \bibinfo{author}{\bibfnamefont{R.}~\bibnamefont{Ruffini}}, \bibnamefont{and}
  \bibinfo{author}{\bibfnamefont{S.-S.} \bibnamefont{Xue}},
  \bibinfo{journal}{Phys. Rev. D} \textbf{\bibinfo{volume}{84}},
  \bibinfo{pages}{084007} (\bibinfo{year}{2011}{\natexlab{b}}).

\bibitem[{\citenamefont{de~Carvalho et~al.}(2014)\citenamefont{de~Carvalho,
  Rotondo, Rueda, and Ruffini}}]{rotondo:PhysRevC.89.015801}
\bibinfo{author}{\bibfnamefont{S.~M.} \bibnamefont{de~Carvalho}},
  \bibinfo{author}{\bibfnamefont{M.}~\bibnamefont{Rotondo}},
  \bibinfo{author}{\bibfnamefont{J.~A.} \bibnamefont{Rueda}}, \bibnamefont{and}
  \bibinfo{author}{\bibfnamefont{R.}~\bibnamefont{Ruffini}},
  \bibinfo{journal}{Phys. Rev. C} \textbf{\bibinfo{volume}{89}},
  \bibinfo{pages}{015801} (\bibinfo{year}{2014}).

\bibitem[{\citenamefont{Kovetz and Shaviv}(1970)}]{kovetz1970thermodynamics}
\bibinfo{author}{\bibfnamefont{A.}~\bibnamefont{Kovetz}} \bibnamefont{and}
  \bibinfo{author}{\bibfnamefont{G.}~\bibnamefont{Shaviv}},
  \bibinfo{journal}{Astronomy and Astrophysics} \textbf{\bibinfo{volume}{8}},
  \bibinfo{pages}{398} (\bibinfo{year}{1970}).

\bibitem[{\citenamefont{Shaviv and Kovetz}(1972)}]{shaviv1972thermodynamics}
\bibinfo{author}{\bibfnamefont{G.}~\bibnamefont{Shaviv}} \bibnamefont{and}
  \bibinfo{author}{\bibfnamefont{A.}~\bibnamefont{Kovetz}},
  \bibinfo{journal}{Astronomy and Astrophysics} \textbf{\bibinfo{volume}{16}},
  \bibinfo{pages}{72} (\bibinfo{year}{1972}).

\bibitem[{\citenamefont{Fantoni}(2017)}]{Fantoni:2017mfs}
\bibinfo{author}{\bibfnamefont{R.}~\bibnamefont{Fantoni}}, \bibinfo{journal}{J.
  Stat. Mech.} \textbf{\bibinfo{volume}{1711}}, \bibinfo{pages}{113101}
  (\bibinfo{year}{2017}), \eprint{arXiv:1709.06064}.

  
\bibitem[{\citenamefont{Han and Fan}(2018)}]{Han:2017kre}
\bibinfo{author}{\bibfnamefont{W.-B.} \bibnamefont{Han}} \bibnamefont{and}
  \bibinfo{author}{\bibfnamefont{X.-L.} \bibnamefont{Fan}},
  \bibinfo{journal}{Astrophys. J.} \textbf{\bibinfo{volume}{856}},
  \bibinfo{pages}{82} (\bibinfo{year}{2018}), \eprint{arXiv:1711.08628}.
  
\bibitem[{\citenamefont{Matsubara}(1955)}]{matsubara1955new}
\bibinfo{author}{\bibfnamefont{T.}~\bibnamefont{Matsubara}},
  \bibinfo{journal}{Progress of theoretical physics}
  \textbf{\bibinfo{volume}{14}}, \bibinfo{pages}{351} (\bibinfo{year}{1955}).

\bibitem[{\citenamefont{Akhiezer and Peletminskii}(1960)}]{akhiezer1960use}
\bibinfo{author}{\bibfnamefont{I.}~\bibnamefont{Akhiezer}} \bibnamefont{and}
  \bibinfo{author}{\bibfnamefont{S.}~\bibnamefont{Peletminskii}},
  \bibinfo{journal}{Zh. Eksp. Teor. Fiz.} \textbf{\bibinfo{volume}{11}},
  \bibinfo{pages}{1316} (\bibinfo{year}{1960}).

\bibitem[{\citenamefont{Freedman and McLerran}(1977)}]{PhysRevD.16.1147}
\bibinfo{author}{\bibfnamefont{B.~A.} \bibnamefont{Freedman}} \bibnamefont{and}
  \bibinfo{author}{\bibfnamefont{L.~D.} \bibnamefont{McLerran}},
  \bibinfo{journal}{Phys. Rev. D} \textbf{\bibinfo{volume}{16}},
  \bibinfo{pages}{1147} (\bibinfo{year}{1977}).  

\bibitem[{\citenamefont{Joyce et~al.}(2018)\citenamefont{Joyce, Barstow,
  Holberg, Bond, Casewell, and Burleigh}}]{joyce:2018mnras481}
\bibinfo{author}{\bibfnamefont{S.~R.~G.} \bibnamefont{Joyce}},
  \bibinfo{author}{\bibfnamefont{M.~A.} \bibnamefont{Barstow}},
  \bibinfo{author}{\bibfnamefont{J.~B.} \bibnamefont{Holberg}},
  \bibinfo{author}{\bibfnamefont{H.~E.} \bibnamefont{Bond}},
  \bibinfo{author}{\bibfnamefont{S.~L.} \bibnamefont{Casewell}},
  \bibnamefont{and} \bibinfo{author}{\bibfnamefont{M.~R.}
  \bibnamefont{Burleigh}}, \bibinfo{journal}{Monthly Notices of the Royal
  Astronomical Society} \textbf{\bibinfo{volume}{481}}, \bibinfo{pages}{2361}
  (\bibinfo{year}{2018}), \eprint{arXiv:1809.01240}.

\bibitem[{\citenamefont{Joseph I.~Kapusta}(2006)}]{book:kapusta}
\bibinfo{author}{\bibfnamefont{C.~G.} \bibnamefont{Joseph I.~Kapusta}},
  \emph{\bibinfo{title}{Finite-Temperature Field Theory: Principles and
  Applications}}, Cambridge Monographs on Mathematical Physics
  (\bibinfo{publisher}{Cambridge University Press}, \bibinfo{year}{2006}),
  \bibinfo{edition}{2nd} ed.

\bibitem[{\citenamefont{Weinberg}(1995)}]{book:17045}
\bibinfo{author}{\bibfnamefont{S.}~\bibnamefont{Weinberg}},
  \emph{\bibinfo{title}{Quantum theory of fields. Foundations}}, vol.
  \bibinfo{volume}{Volume 1} (\bibinfo{publisher}{Cambridge University Press},
  \bibinfo{year}{1995}), \bibinfo{edition}{1st} ed.

\bibitem[{\citenamefont{Nair}(2005)}]{book:16435}
\bibinfo{author}{\bibfnamefont{V.~P.} \bibnamefont{Nair}},
  \emph{\bibinfo{title}{Quantum Field Theory: A Modern Perspective}}, Graduate
  Texts in Contemporary Physics (\bibinfo{publisher}{Springer},
  \bibinfo{year}{2005}), \bibinfo{edition}{1st} ed.

\bibitem[{\citenamefont{Hossain and Mandal}(2021)}]{hossain2021equation}
\bibinfo{author}{\bibfnamefont{G.~M.} \bibnamefont{Hossain}} \bibnamefont{and}
  \bibinfo{author}{\bibfnamefont{S.}~\bibnamefont{Mandal}},
  \bibinfo{journal}{Journal of Cosmology and Astroparticle Physics}
  \textbf{\bibinfo{volume}{2021}}, \bibinfo{pages}{026} (\bibinfo{year}{2021}).

\bibitem[{Hossain and Mandal(2021)}]{hossain2021higher}
Hossain GM, Mandal S (2021) Higher mass limits of neutron stars from the
  equation of states in curved spacetime. arXiv:2109.09606  

\bibitem[{\citenamefont{Hossain and Mandal}(2019)}]{Hossain:2019cml}
\bibinfo{author}{\bibfnamefont{G.~M.} \bibnamefont{Hossain}} \bibnamefont{and}
  \bibinfo{author}{\bibfnamefont{S.}~\bibnamefont{Mandal}}
  (\bibinfo{year}{2019}), \eprint{arXiv:1904.09779}.

\bibitem[{\citenamefont{Linares et~al.}(2018)\citenamefont{Linares, Shahbaz,
  and Casares}}]{linares2018peering}
\bibinfo{author}{\bibfnamefont{M.}~\bibnamefont{Linares}},
  \bibinfo{author}{\bibfnamefont{T.}~\bibnamefont{Shahbaz}}, \bibnamefont{and}
  \bibinfo{author}{\bibfnamefont{J.}~\bibnamefont{Casares}},
  \bibinfo{journal}{The Astrophysical Journal} \textbf{\bibinfo{volume}{859}},
  \bibinfo{pages}{54} (\bibinfo{year}{2018}).

\bibitem[{\citenamefont{Cromartie et~al.}(2020)\citenamefont{Cromartie,
  Fonseca, Ransom, Demorest, Arzoumanian, Blumer, Brook, DeCesar, Dolch, Ellis
  et~al.}}]{cromartie2020relativistic}
\bibinfo{author}{\bibfnamefont{H.~T.} \bibnamefont{Cromartie}},
  \bibinfo{author}{\bibfnamefont{E.}~\bibnamefont{Fonseca}},
  \bibinfo{author}{\bibfnamefont{S.~M.} \bibnamefont{Ransom}},
  \bibinfo{author}{\bibfnamefont{P.~B.} \bibnamefont{Demorest}},
  \bibinfo{author}{\bibfnamefont{Z.}~\bibnamefont{Arzoumanian}},
  \bibinfo{author}{\bibfnamefont{H.}~\bibnamefont{Blumer}},
  \bibinfo{author}{\bibfnamefont{P.~R.} \bibnamefont{Brook}},
  \bibinfo{author}{\bibfnamefont{M.~E.} \bibnamefont{DeCesar}},
  \bibinfo{author}{\bibfnamefont{T.}~\bibnamefont{Dolch}},
  \bibinfo{author}{\bibfnamefont{J.~A.} \bibnamefont{Ellis}},
  \bibnamefont{et~al.}, \bibinfo{journal}{Nature Astronomy}
  \textbf{\bibinfo{volume}{4}}, \bibinfo{pages}{72} (\bibinfo{year}{2020}).

\bibitem[{\citenamefont{Serot}(1992)}]{serot1992quantum}
\bibinfo{author}{\bibfnamefont{B.~D.} \bibnamefont{Serot}},
  \bibinfo{journal}{Reports on Progress in Physics}
  \textbf{\bibinfo{volume}{55}}, \bibinfo{pages}{1855} (\bibinfo{year}{1992}).
  
\bibitem[{\citenamefont{Serot and Walecka}(1997)}]{serot1997recent}
\bibinfo{author}{\bibfnamefont{B.~D.} \bibnamefont{Serot}} \bibnamefont{and}
  \bibinfo{author}{\bibfnamefont{J.~D.} \bibnamefont{Walecka}},
  \bibinfo{journal}{International Journal of Modern Physics E}
  \textbf{\bibinfo{volume}{6}}, \bibinfo{pages}{515} (\bibinfo{year}{1997}).

\bibitem[{\citenamefont{Whittenbury et~al.}(2014)\citenamefont{Whittenbury,
  Carroll, Thomas, Tsushima, and Stone}}]{whittenbury2014quark}
\bibinfo{author}{\bibfnamefont{D.}~\bibnamefont{Whittenbury}},
  \bibinfo{author}{\bibfnamefont{J.}~\bibnamefont{Carroll}},
  \bibinfo{author}{\bibfnamefont{A.}~\bibnamefont{Thomas}},
  \bibinfo{author}{\bibfnamefont{K.}~\bibnamefont{Tsushima}}, \bibnamefont{and}
  \bibinfo{author}{\bibfnamefont{J.}~\bibnamefont{Stone}},
  \bibinfo{journal}{Physical Review C} \textbf{\bibinfo{volume}{89}},
  \bibinfo{pages}{065801} (\bibinfo{year}{2014}).

\bibitem[{\citenamefont{Katayama et~al.}(2012)\citenamefont{Katayama, Miyatsu,
  and Saito}}]{katayama2012equation}
\bibinfo{author}{\bibfnamefont{T.}~\bibnamefont{Katayama}},
  \bibinfo{author}{\bibfnamefont{T.}~\bibnamefont{Miyatsu}}, \bibnamefont{and}
  \bibinfo{author}{\bibfnamefont{K.}~\bibnamefont{Saito}},
  \bibinfo{journal}{The Astrophysical Journal Supplement Series}
  \textbf{\bibinfo{volume}{203}}, \bibinfo{pages}{22} (\bibinfo{year}{2012}).

\bibitem[{\citenamefont{Miyatsu et~al.}(2013)\citenamefont{Miyatsu, Yamamuro,
  and Nakazato}}]{miyatsu2013new}
\bibinfo{author}{\bibfnamefont{T.}~\bibnamefont{Miyatsu}},
  \bibinfo{author}{\bibfnamefont{S.}~\bibnamefont{Yamamuro}}, \bibnamefont{and}
  \bibinfo{author}{\bibfnamefont{K.}~\bibnamefont{Nakazato}},
  \bibinfo{journal}{The Astrophysical Journal} \textbf{\bibinfo{volume}{777}},
  \bibinfo{pages}{4} (\bibinfo{year}{2013}).
      
\bibitem[{Zhang and Li(2019)}]{zhang2019extracting}
Zhang NB, Li BA (2019) Extracting nuclear symmetry energies at high densities
  from observations of neutron stars and gravitational waves. The European
  Physical Journal A 55(3):1--23

\bibitem[{Lackey et~al.(2006)Lackey, Nayyar, and
  Owen}]{lackey2006observational}
Lackey BD, Nayyar M, Owen BJ (2006) Observational constraints on hyperons in
  neutron stars. Physical Review D 73(2):024021

\bibitem[{Takami et~al.(2014)Takami, Rezzolla, and
  Baiotti}]{takami2014constraining}
Takami K, Rezzolla L, Baiotti L (2014) Constraining the equation of state of
  neutron stars from binary mergers. Physical Review Letters 113(9):091104

\bibitem[{Chatziioannou(2020)}]{chatziioannou2020neutron}
Chatziioannou K (2020) Neutron-star tidal deformability and equation-of-state
  constraints. General Relativity and Gravitation 52(11):1--49
    
\end{thebibliography}

\end{document}